\documentclass[usenatbib,useAMS]{mnras}
\usepackage{newtxtext,newtxmath}
\usepackage[T1]{fontenc}
\usepackage{ae,aecompl}
\usepackage{graphicx} % Including figure files
\usepackage{amsmath,amsfonts}  % Advanced maths commands
\usepackage{widetext} % Wide equations
\usepackage{xparse,xspace}
\usepackage{cancel}
\usepackage{siunitx}
\usepackage{placeins}
\usepackage{caption}

\defcitealias{Finley2017}{FM17}
\defcitealias{MurrayClay2009}{MC09}

\DeclareSIUnit\ergs{ergs}
\DeclareSIUnit\gauss{G}
\DeclareSIUnit\msun{M_\odot}
\DeclareSIUnit\lsun{L_{\odot}}
\DeclareSIUnit\rsun{R_{\odot}}

\newcommand{\mast}{M_{\ast}}
\newcommand{\rast}{R_{\ast}}
\newcommand{\athena}{\texttt{Athena++}\xspace} % athena++
\newcommand{\surf}{\mathcal{S}} % Calligraphy S
\newcommand{\rproc}{\ensuremath{r}-process\xspace}
\newcommand{\esc}{\mathrm{esc}}

\newcommand{\uvec}[1]{\boldsymbol{\hat{#1}}} % Unit Vector
\newcommand{\vecomega}{\boldsymbol{\varOmega}} % bold italic omega
\NewDocumentCommand\pder{mmg}{\ensuremath{
    \IfNoValueTF{#3}
    {\dfrac{\partial #1}{\partial #2}}
    {\left(\dfrac{\partial #1}{\partial #2}\right)_{#3}}
}}

\newcommand\tstrut{\rule{0pt}{2.6ex}}         % = `top' strut
\newcommand\bstrut{\rule[-0.9ex]{0pt}{0pt}}   % = `bottom' strut

\newcommand\thefont{\expandafter\string\the\font}

\title[Magnetized Winds]{Magnetized Rotating Isothermal Winds}
\pubyear{2022}
\author[M. J. Raives et al.]{
Matthias J. Raives$^{1,2,3}$\thanks{E-mail: \href{mailto:mraives@carnegiescience.edu}{mraives@carnegiescience.edu}},
Matthew S.\ B.\ Coleman$^{4,5}$,
\& Todd A. Thompson$^{2,3,6}$
\\
${}^{1}$The Observatories of the Carnegie Institution for Science, 813 Santa Barbara St., Pasadena, CA 91101, USA;\\
${}^{2}$Department of Astronomy, The Ohio State University, Columbus, OH 43210, USA\\
${}^{3}$Center for Cosmology and Astroparticle Physics, The Ohio State University, Columbus, OH 43210, USA\\
${}^{4}$Department of Astrophysical Sciences, 4 Ivy Lane, Princeton University, Princeton, NJ 08540, USA\\
${}^{5}$Department of Physics and Engineering Physics, Stevens Institute of Technology,
Castle Point on the Hudson, Hoboken, NJ 07030, USA\\
${}^{6}$Department of Physics, The Ohio State University, Columbus, OH 43210, USA\\
}

\begin{document}

    \label{firstpage}
    \pagerange{\pageref{firstpage}--\pageref{lastpage}}
    \maketitle
    
    \begin{abstract}
        We consider the general problem of a Parker-type non-relativistic isothermal wind from a rotating and magnetic star.  Using the magnetohydrodynamics (MHD) code \athena, we construct an array of simulations in the stellar rotation rate $\Omega_\ast$ and the isothermal sound speed $c_T$, and calculate the mass, angular momentum, and energy loss rates across this parameter space.  We also briefly consider the three dimensional case, with misaligned magnetic and rotation axes.
        We discuss applications of our results to the spindown of normal stars, highly-irradiated exoplanets, and to nascent highly-magnetic and rapidly-rotating neutron stars born in massive star core collapse.
    \end{abstract}
    
    
    \section{Introduction}
        
        The spindown of stars\footnote{For simplicity, we use ``star'' throughout this paper to refer to a generic central body.} as a result of the torque caused by their thermal magnetocentrifugal winds is a classic topic with connections throughout astrophysics. Strong magnetic fields force the wind to effectively co-rotate with the star out to the Alfv\'en radius $R_A$, where the poloidal kinetic energy density is approximately equal to the magnetic energy density.  By forcing the wind to co-rotate with the star out to large radii, thermal magnetocentrifugal winds can quickly shed large amounts of angular momentum, resulting in rapid spindown.
        
        This problem goes back at least to \citet{Schatzman1962} and \citet{Weber1967}, who investigated the physics of a spherically symmetric magnetized wind, following the classic work by \cite{Parker1958}. \citet{Sakurai1985} and \citet{Mestel1987} later investigated the problem in two dimensions. Magnetocentrifugal braking is invoked in many astrophysical contexts to explain high energy winds and rapid rotational braking of not only stars, but also disks (e.g.,  \citealt{Blandford1982}).
        %, and exoplanets, (e.g., \citealt{Owen2014}).
        This problem has been investigated most thoroughly in the context of stellar winds, both thermal \citep{Keppens2000,Vidotto2014} and line-driven \citep{Doula2002}.  3D magnetohydrodynamic (MHD) simulations of the Sun's corona and the solar wind go back decades \citep{Suess1977,Suess1999,Lionello2001,Lionello2009,Gressl2014,Reiss2016}, though most such models are calibrated to the specific conditions of the Sun.  \citet{Finley2017,Finley2018} have also investigated this problem in the case of magnetized winds with higher-order multipole structure.  Magnetocentrifugal braking is also important in discussions of gyrochronology \citep{Saders2016}, and magnetocentrifugal winds have also been invoked in discussion of hot Jupiter atmospheres \citep{Owen2014}.
        
        The discussion of stellar spindown also intersects with discussion of proto-neutron star (PNS) and proto-magnetar winds, and the magnetocentrifugal explosion mechanism for core-collapse supernovae (CCSNe).  In a core-collapse supernova, after the shockwave is revived and driven outwards into the surrounding massive star progenitor, the cooling PNS core drives a wind into the post-supernova-shock environment \citep{Woosley1994,Janka1995,Burrows1995}.  The physics of thermal and purely hydrodynamic (i.e., non-magnetized) winds from PNSs was first explored in \citet{Duncan1986}, who considered them in spherical symmetry (1D).
        %, pure hydrodynamic winds with a neutrino-coupled equation of state (EOS), but who did not include any magnetocentrifugal effects.  
        Interest in this phase of evolution was piqued by \citet{Woosley1994}, who found that their thermodynamic conditions might be conducive to the \rproc, though later studies \citep{Qian1996,Hoffman1997,Otsuki2000,Thompson2001,Wanajo2001} showed that normal PNS winds do not enter a thermodynamic regime conducive to production of the heaviest \rproc elements unless they are strongly magnetized (see \citealt{Thompson2003b, Thompson2018}).  Later 1D studies \citep{Thompson2004,Metzger2007} considered the effects of magnetic fields on PNS winds, and showed that magnetar strength ($B_{\ast}\sim\SI{e15}{\gauss}$) magnetic fields would dynamically dominate the winds on second-long timescales after explosion \citep{CThompson2001}, driving them to energies comparable to gamma-ray bursts (GRBs) and superluminous supernovae (SLSNe).  The connection to GRBs has been further explored in \citet{Thompson2004,Thompson2007,Bucciantini2009,Metzger2011}, and the SLSNe connection in \citet{Wheeler2000,Komissarov2007,Kasen2010,Woosley2010,Dong2016,Chatzopoulos2016}.  Some models also link magnetocentrifugal braking to ordinary (i.e., not superluminous) CCSNe \citep{Ostriker1971,Symbalisty1984,Sukhbold2017}, in addition to other magnetohydrodynamic effects, such as the magnetorotational instability \citep{Akiyama2003,Thompson2005,Nishimura2017}.
        
        Despite the broad reach of this problem, there are still key insights missing from the literature.  First, most studies of this problem restrict to a narrow range of parameter space.  Second, while 2D simulations of this magnetocentrifugal winds are becoming more common \citep{Bucciantini2006,Finley2017}, 3D simulations are still relatively new (e.g., \citealt{Vurm2021}, but see, e.g., \citealt{Vidotto2009} in the stellar context), and few of them model the effects of magnetospheres not aligned with the axis of rotation (e.g., \citealt{Subramanian2022}).  This paper is an effort to fill in these gaps.  We present MHD simulations of 2D and 3D, non-relativistic, magnetocentrifugal winds, using an isothermal equation of state (EOS).  While the isothermal EOS is limiting, it also allows us to simultaneously compare these results to existing results in the literature of stellar winds and highly-irradiated ``hot Jupiter" winds, in addition to proto-magnetar winds.  
        
        In \S\ref{sec:Methods}, we present details of our simulations, including the initial and boundary conditions used, as well as the details of the rotating reference frame.  In \S\ref{sec:Results}, we present our findings: how the eigenvalues of the problem -- the mass loss rate $\dot{M}$ and the angular momentum loss rate $\dot{J}$ -- scale with the various simulation parameters.  In particular, we construct a parameter space covering a wide range of rotation rates and sound speeds
        %\footnote{The isothermal sound speed can be taken as a measure of temperature, and thus a proxy for the energy source of the system, e,g., the UV flux for hot Jupiter winds, or the neutrino luminosity (and therefore, given a known luminosity evolution such as given by \citealt{Pons1999}, the time since birth) for a proto-magnetar.} 
        at a representative magnetic field strength.  In \S\ref{sec:Discussion}, we consider our results in the context of proto-magnetar winds, as well as in the context of sun-like stars and irradiated hot Jupiters.  Finally, in \S\ref{sec:Conclusions}, we discuss our findings in the context of the field and identify areas for future investigation.
    
    \section{Methods}\label{sec:Methods}
        We perform MHD simulations in \athena{} \citep{Stone2019}, which we have configured to solve the isothermal MHD equations:
        \begin{align}
          \pder{\rho}{t}+\nabla\cdot\left(\rho\mathbfit{v}\right)&=0\label{eq:continuity}\\
          \pder{\rho\mathbfit{v}}{t}+\nabla\cdot\left[\rho\mathbfit{vv}+\left(\rho c_T^2 +\dfrac{B^2}{2}\right)\mathbf{I}-\mathbfit{BB}\right]&=-\rho\nabla\varphi\label{eq:momentum}\\
          %\pder{E}{t}+\nabla\cdot\left[\left(E+p\right)\mathbfit{v}-\mathbfit{B}\left(\mathbfit{B}\cdot\mathbfit{v}\right)\right]&=0\label{eq:energy}\\
          \pder{\mathbfit{B}}{t}-\nabla\times\left(\mathbfit{v}\times\mathbfit{B}\right)&=0,\label{eq:eulermag}
        \end{align}
        where $\varphi$ is the gravitational potential, which we assume to be the point mass potential of the star:
        \begin{equation}
            \varphi = \frac{G\mast}{r},
        \end{equation}
        and $c_T$ is the isothermal sound speed.
        
        \subsection{Rotating Reference Frame}
            %In order to simplify the magnetic field boundary conditions at the inner boundary, 
            We perform our simulations in the rotating reference frame, i.e., a reference frame where
            \begin{equation}
                \phi\to\phi' + \varOmega_{\ast} t'
            \end{equation}
            and
            \begin{equation}
                \mathbfit{v} \to \mathbfit{v}' + \vecomega_{\ast}\times\mathbfit{r}',
            \end{equation}
            where $\vecomega_{\ast}=\varOmega_{\ast}\uvec{z}$ is the angular velocity of the frame.  By construction, this is equal to the (inertial frame) angular velocity of the fluid at $r=\rast$ (see the following section).
            
            Here (and throughout), primed quantities refer to values measured in the rotating reference frame, and unprimed quantities refer to values measured in the inertial (lab) frame.  Though $r$, $\theta$ and $t$ do not change between reference frames, we still denote them with primes when measuring quantities in the rotating frame, for the sake of clarity and accuracy.
            
            Since the rotating reference frame is non-inertial, the momentum equation (Equation~\ref{eq:momentum}) must be changed to account for the pseudo-forces arising from measurement in this frame.  These forces are the familiar Coriolis and centrifugal forces (since $\boldsymbol{\dot{\varOmega}}_{\ast}=0$, the Euler force is zero):
            \begin{align}
                \mathbfit{a}_\mathrm{cor}' &= -2\vecomega_{\ast} \times \mathbfit{v}'\\
                \mathbfit{a}_\mathrm{cen}' &= -\vecomega_{\ast} \times (\vecomega_{\ast} \times \mathbfit{r}').
            \end{align}
            
            We discuss the mathematics of the rotating reference frame, and its implementation in \athena, in more detail in Appendix~\ref{app:rotatingframe}.
        
        \subsection{Initial Conditions}
          
            We initialize all simulations with the density field in spherically-symmetric hydrostatic equilibrium, with the central density, core mass, and core radius normalized to
            \begin{align}
                \rho_{\ast} &= 1\label{eq:coderho}\\
                G\mast &= 1\label{eq:codeGM}\\
                \rast &= 1.\label{eq:codeR}
            \end{align}
            We initialize the velocity field as a pure radial field plus an angular-momentum conserving angular velocity field. The initial radial velocity is chosen such that it increases slowly at large radii and has a finite value at $r=\rast$.  Our chosen velocity field relaxes to the ``correct'' radial velocity field during an early, transient phase of the simulation.  The initial velocity field is
            \begin{equation}
                \mathbfit{v} = 2{c_T}\left(\frac{r}{r_{s}}\right)^{{1/2}}\uvec{r} + \vecomega\times\mathbfit{r},
            \end{equation}
            where $\vecomega$ is the angular velocity profile
            \begin{equation}
                \vecomega = \varOmega_{\ast}\left(\frac{\rast}{r}\right)^2\uvec{z},
            \end{equation}
            $r_s$ is the sonic radius (for a non-rotating, non-magnetic isothermal wind)
            \begin{equation}
                r_{s} = \frac{GM}{2c_{T}^{2}},\label{eq:rsonic}
            \end{equation}
            and $\varOmega_{\ast}$ is the angular velocity at $r=\rast$.  In the rotating reference frame, this becomes
            \begin{equation}
                \mathbfit{v}' = 2{c_T}\left(\frac{r'}{r_{s}}\right)^{{1/2}}\uvec{r}' + (\vecomega-\vecomega_{\ast})\times\mathbfit{r}'.
            \end{equation}
            We note that our results are not very sensitive to the initial radial velocity profile, so long as the sonic point is on the grid (i.e., $v_r(\rast)<c_T$ and $v_r(R_\mathrm{max})>c_T$).
            
            We initialize the magnetic field as a dipole field, specified by the vector potential
            \begin{equation}
                \mathbfit{A} = \frac{B_{\ast}}{2}\frac{\rast^3}{r^2}\left(\begin{array}{c}
                     0\\-\sin\alpha\sin\phi\\\cos\alpha\sin\theta - \cos\theta\cos\phi\sin\alpha
                \end{array}\right),
            \end{equation}
            where $B_{\ast}$ is the magnetic field strength at the pole at ${r=\rast}$ and $\alpha$ is the angle between the the rotational axis and magnetic axis.  That is,
            \begin{equation}
                \alpha\equiv\arccos{\left(\frac{\mathbfit{m}}{m}\cdot\frac{\vecomega_{\ast}}{\varOmega_{\ast}}\right)},
            \end{equation}
            where $\mathbfit{m}$ is the magnetic dipole moment.  The plane defined by these two vectors is the $\phi=0$ plane.  The tilted dipole is discussed in more detail in \S\ref{sec:tilt} and Appendix~\ref{app:tilt}.   We also note that, because our calculations are non-relativistic, we can assume $\mathbfit{B}'=\mathbfit{B}$.
            
            Thus, there are four free parameters we must specify: the sound speed $c_{T}$, the tilt angle $\alpha$, the polar magnetic field strength $B_{\ast}$, and the core angular velocity $\varOmega_{\ast}$.  The field strength is set by the dimensionless parameter
            \begin{equation}
                \xi_{B}\equiv \left.\frac{v_{A}}{v_{\mathrm{esc}}}\right|_{\rast} = \left({\frac{B_{\ast}^{2}\rast}{8\upi G\mast\rho_{\ast}}}\right)^{1/2},\label{eq:mag}
            \end{equation}
            the core angular velocity is specified by the dimensionless parameter
            \begin{equation}
                \xi_{\varOmega}\equiv \left.\frac{v_{\phi}}{v_{\mathrm{esc}}}\right|_{\rast} = \left({\frac{\varOmega_{\ast}^{2}\rast^{3}}{2G\mast}}\right)^{1/2},\label{eq:rot}
            \end{equation}
            and the sound speed is specified by the dimensionless parameter
            \begin{equation}
                \xi_T\equiv \left.\frac{c_T}{v_\esc}\right|_{\rast} = \left(\frac{c_T^2\rast}{2GM}\right)^{1/2}.\label{eq:therm}
            \end{equation}

            By using these dimensionless parameters, and by using our dimensionless ``code units'' (Equations~\ref{eq:coderho}-\ref{eq:codeR}), we can more readily scale the problem to a number of other contexts.  This is discussed in more detail in Appendix~\ref{app:units}.
        
        \subsection{Boundary Conditions}
            At the inner boundary, we enforce boundary conditions such that $\frac{\partial\mathbfit{v}}{\partial r}=0,$  $\frac{\partial\mathbfit{B}'}{\partial r}=0,$ and $v_r\ge 0$ in the ghost zones. We also set the density to preserve (non-rotating) hydrostatic equilibrium across the ghost zones:
            \begin{equation}
                \rho(r) = %\rho_{\ast}\exp\left[\frac{GM}{c_{T}^{2}}\left(\frac{1}{r}-\frac{1}{\rast}\right)\right].
                \rho_\ast \exp\left[\frac{1}{2\xi_T^2}\left(\frac{\rast}{r} - 1\right)\right].
            \end{equation}
            In order to enforce
            \begin{equation}
                \frac{\partial B_r'}{\partial t'} = 0
            \end{equation}
            at the inner boundary (and thus prevent unchecked growth in $\mathbfit{B}'$), we fix the components of the electric field $\mathbfit{E}'$ at the inner boundary:
            \begin{align}
                E_\theta' &= 0\\
                E_\phi' &= 0.
            \end{align}
            At the outer boundary, we enforce ``outflow'' boundary conditions such that:
            \begin{align}
                \rho^{N_1+i} &= \left(\frac{r^{N_1}}{r^{N_1+i}}\right)^2\rho^{N_1}\\
                v_r^{N_1+i} &= v_r^{N_1}\\
                v_\theta^{N_1+i} &= \frac{r^{N_1+i}}{r^{N_1}}v_\theta^{N_1}\\
                (v_\phi')^{N_1+i} &= \frac{r^{N_1+i}}{r^{N_1}}(v_\phi')^{N_1}\\
                \mathbfit{B}^{N_1+i} &= \mathbfit{B}^{N_1},
            \end{align}
            where $N_1$ is the index of the last active radial zone.  The boundary conditions are discussed in greater detail in Appendix~\ref{app:bc}.
            
        \subsection{Resolution and Grid}\label{sec:Methods_Resolution}
            The resolution and box size we use varies depending on the simulation.  By convention, we use a $1024\times1\times1$ grid in 1D (spherical symmetry), a $512\times512\times1$ grid in 2D (axisymmetry), and a $512\times64\times128$ grid in full 3D, with a box that has physical extent $r\in[1,50]\rast$, $\theta\in[0,\upi]$, $\phi\in[0,2\upi]$ (except in 1D, where the angular extent of the box is reduced to $\theta\in\left[\frac{\upi}{2}-\frac{1}{2},\frac{\upi}{2}+\frac{1}{2}\right]$, $\phi\in[-0.5,0.5]$).  Zones in the $\theta$ and $\phi$ directions are linearly spaced, while zones in the radial direction are logarithmically spaced.
            
            Some simulations require a larger outer boundary to fully capture the sonic surfaces (see \S\ref{sec:structure}).  For these simulations, we expand the outer boundary to $100\rast$.  The number of radial zones used in these simulations is not altered, as, due to the logarithmic spacing in the radial direction, this increase in box size does not lead to a significant reduction in resolution.
            
        \subsection{Simulation Stop Time}
            All of our simulations are evolved until a steady-state behavior emerges.  For our 2D simulations, we simulate until $t=1000$ in our code units (Equations \eqref{eq:coderho} - \eqref{eq:codeR}).  However, this is well in excess of the time required for steady state behavior to emerge.  For our 3D simulations, we end the simulations early.  The time varies between simulations but is in general $t\gtrsim200$, again in our code units.
            
            %We discuss the effect of different resolutions on our results in \S\ref{sec:Methods_Resolution}.
    
    \section{Results}\label{sec:Results}
    
        \subsection{Non-Rotating, Non-Magnetic Baseline}
\begin{figure}
                \centering
                \includegraphics[width=\linewidth]{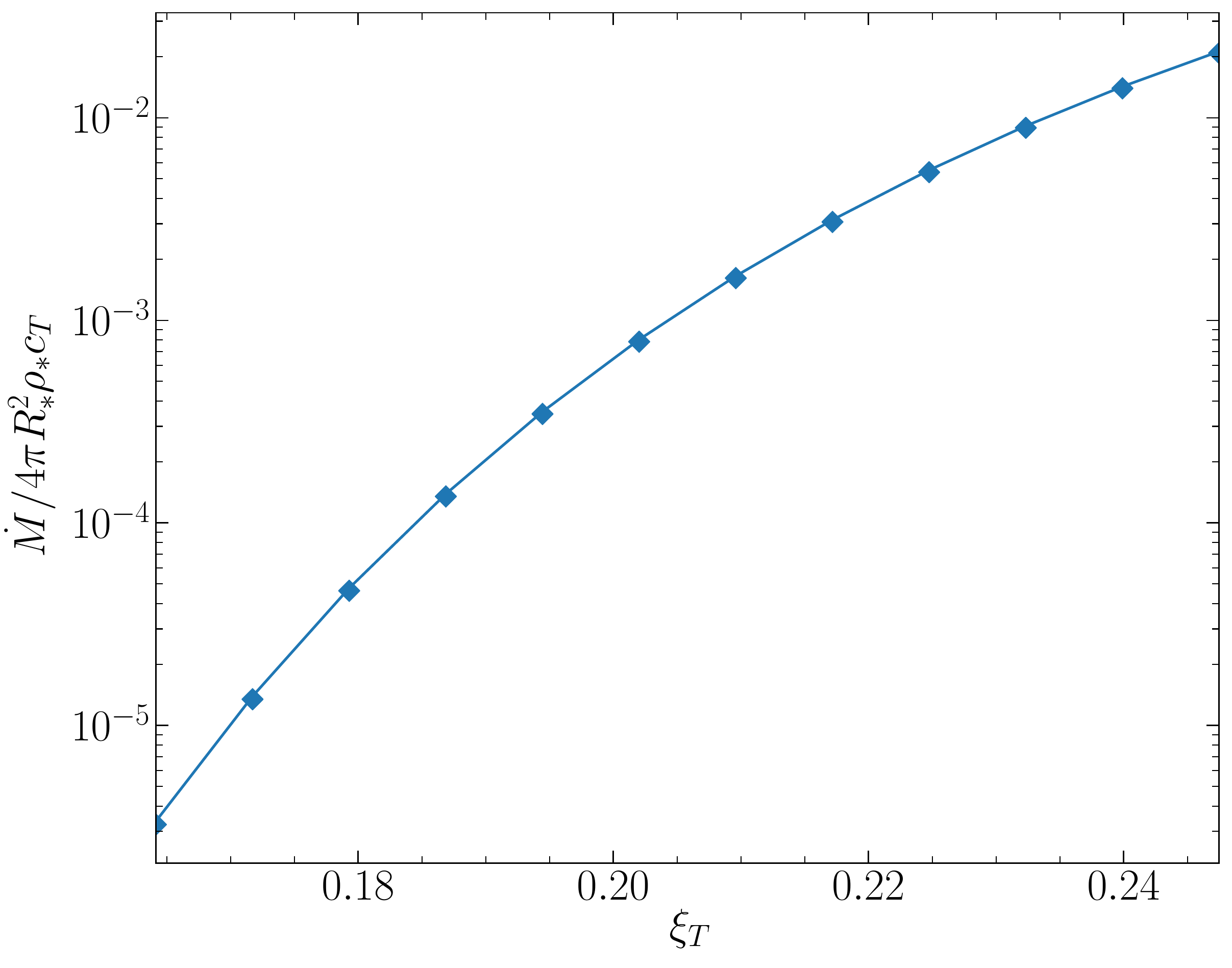}
                \caption{Normalized mass loss rate versus the isothermal sound speed (blue points), and the analytic solution (Equation~\ref{eq:mdot-an}) for a pure isothermal wind (orange curve).}
                \label{fig:nrnm}
            \end{figure}

            We performed high resolution simulations of 1D, non-rotating, non-magnetic (NRNM) Parker winds to serve as a baseline comparison for the rotating and magnetic simulations.  In Figure~\ref{fig:nrnm}, we plot the normalized mass loss rate as a function of the isothermal sound speed.  We find that the mass loss rapidly decreases as $c_T$ decreases, falling by several orders of magnitude over a $\sim\!30$ per-cent decrease in $c_T$. Analytically \citep{Lamers1999}, the mass loss rate is
            \begin{equation}
                \dot{M} = \frac{\upi}{4} \rast^2\rho_{\ast} c_T \left(\frac{v_\mathrm{esc}}{c_T}\right)^4\exp\left[\frac{3}{2}-\frac{v_\mathrm{esc}^2}{2c_T^2}\right],\label{eq:mdot-an}
            \end{equation}
            where $v_\mathrm{esc}$ is the escape velocity at $r=\rast$.  We see that our simulations closely follow the analytic solution (to within a few percent at this resolution; see \S\ref{sec:Methods_Resolution} and \S\ref{sec:results_resolution} for more details regarding convergence).
            
        \subsection{Properties of the Magnetized Winds}
        \label{sec:propwinds}
\begin{figure*}
                \centering
                \includegraphics[width=\textwidth]{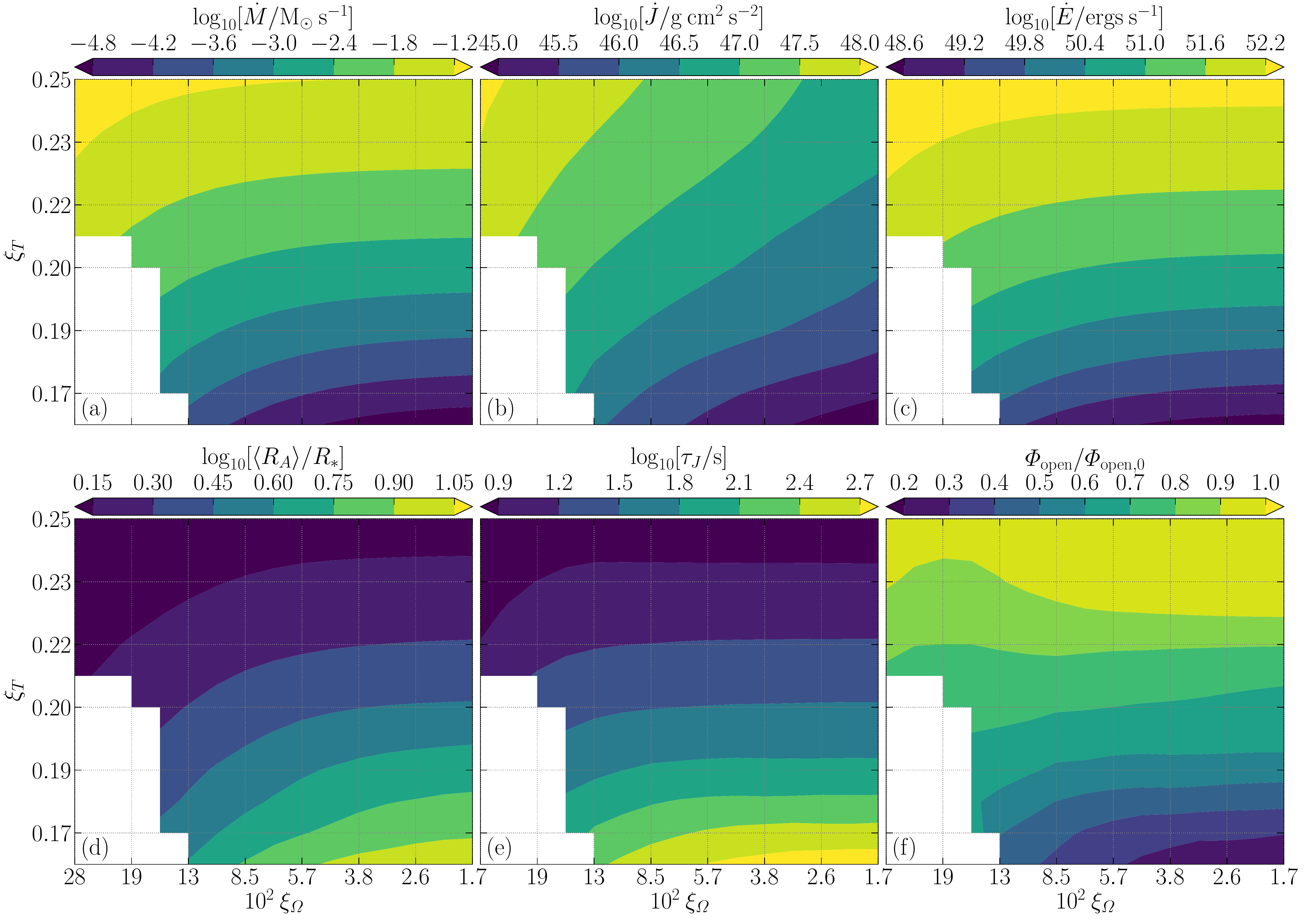}
                \caption{From left to right, top to bottom: $(a)$ the wind mass loss rate, $(b)$ angular momentum loss rate (i.e., wind torque), $(c)$ energy loss rate, $(d)$ average Alfv\'en radius, $(e)$ spindown time, and $(f)$ unsigned open magnetic flux, over a grid in dimensionless isothermal sound speed $\xi_T$ (Equation~\ref{eq:therm}) and dimensionless rotation rate $\xi_{\varOmega}$ (Equation~\ref{eq:rot}), for a magnetic field of $\xi_B=\num{4.63e-2}$.  For dimensional quantities (panels $a,b,c,e$), the units assume parameters are scaled to the proto-magnetar regime (see \S\ref{sec:Discussion} and Appendix~\ref{app:units} for more details).  With the exception of $\dot{J}$, which remains a strong function of $\xi_\varOmega$ even at small rotation rates, these quantities are all strong functions of $\xi_T$ and weak functions of $\xi_\varOmega$ except for the most rapid rotators.}
                \label{fig:heatmap}
            \end{figure*}

            In Figure~\ref{fig:heatmap}, we present the results of a suite of 2D axisymmetric simulations across a parameter space of $0.1414\leq\xi_T\leq0.2475$ and $0.0174\leq\xi_\varOmega\leq0.2777$, for $\xi_B=\num{4.63e-2}$.  The data summarized in this figure is also presented in table form in Appendix~\ref{sec:fulltable}.  We discuss each panel of this figure in the following sections.
            
            In certain cases with low $\xi_{T}$ and high $\xi_{\varOmega}$ (the bottom left portion of the panels in Figure~\ref{fig:heatmap}), we encounter significant numerical errors that prevent us from completing accurate simulations using those parameters.  This region is colored in white in the figure and is not included in Appendix~\ref{sec:fulltable}.
            
            \subsubsection{Mass Loss Rate}
            In the steady-state limit, the continuity equation (Equation~\ref{eq:continuity}) implies that\footnote{$\mathrm{d}\Omega$ here is the solid angle element, $\mathrm{d}\Omega=\sin\theta\:\mathrm{d}\theta\:\mathrm{d}\phi$, and should not be confused with the angular speed $\varOmega$.}
            \begin{equation}
                \dot{M} = \oint r^2\rho v_r\:\mathrm{d}\Omega = \mathrm{constant}.
            \end{equation}
            Unlike in the spherically symmetric case, however, the mass flux is not equally distributed over solid angle; i.e., ${\dot{M}\neq4\upi r^2\rho v_r}$.  We plot the mass loss rate (normalized to PNS values) in panel (a) of Figure~\ref{fig:heatmap}.  As in the NRNM case, we find that the mass loss rate is a strong function of the sound speed, but in addition, that there is a substantial centrifugal enhancement to $\dot{M}$ for large $\xi_\varOmega$, as can be seen by the downturn of the $\dot{M}$ contours.
            
            Specifically, the $\dot{M}$ values shown in Figure~\ref{fig:heatmap} are the mass-weighted radial averages, i.e.,
            \begin{equation}
                \langle \dot{M} \rangle = \frac{\sum_{r} \dot{M} M_{r}}{\sum_{r} M_{r}}
            \end{equation}
            where $M_{r}$ is the sum of cell masses over the spherical surface at radial coordinate $r$:
            \begin{equation}
                M_{r} = \sum_{\theta,\phi}\rho r^{2}\sin\theta\Delta_{r}\Delta_{\theta}\Delta_{\phi},
            \end{equation}
            where $\sum_{k}$ indicates the sum over all cells in the $k$-direction, and $\Delta_{k}$ is the cell width in the $k$-direction.  The sum is taken over all cells with $r>10\rast$ (except the outer 5 cells, which are excluded to ignore boundary effects).  We use a similar method to measure the other quantities shown in Figure~\ref{fig:heatmap}, but in the case of $\varPhi_{\rm open}$ (see below), the average is taken for all $r>30\rast$ to ensure that the average only includes open field lines.
            
            \subsubsection{Angular Momentum Loss Rate}
            Following the derivation in \citet{Vidotto2014}, the $z$-component of the angular momentum flux across a closed spherical surface is given by,
            \begin{equation}
                \dot{J} = \oint r^2 \rho v_r\varpi\left(v_\phi - \frac{B_rB_\phi}{4\upi\rho v_r}\right)\:\mathrm{d}\Omega,\label{eq:jdot}
            \end{equation}
            where $\varpi=r\sin\theta$ is the cylindrical radius.  Because the angular momentum of the magnetar is aligned with the $z$-axis (in both the aligned and tilted cases), this is the only relevant component of the flux.  We plot this quantity in panel (b) of Figure~\ref{fig:heatmap}.
            
            We can also use the angular momentum loss rate to define the spindown timescale:
            \begin{equation}
                \tau_J = \frac{J}{\dot{J}},\label{eq:tauJ}
            \end{equation}
            where $J$ is the angular momentum of the star, which we take to be that of a uniform density spherical star,
            \begin{equation}
                J = \frac{2}{5}MR_{\ast}^2\varOmega_{\ast}.
            \end{equation}
            The spindown timescale is shown in panel (e) of Figure~\ref{fig:heatmap}.
            
            \subsubsection{Energy Loss Rate}
            For an equatorial rotator (i.e., one with $v_\theta=B_\theta=0$ ; e.g., \citealt{Weber1967})\footnote{Though we explicitly include the $v_\theta$ term in our calculations.}, one can show that the quantity
            \begin{equation}
                \mathcal{B}_{\rm eq} \equiv \frac{1}{2}(v_r^2 + v_\theta^2 + v_\phi^2) + c_T^2\log\frac{\rho}{\rho_\ast} - \frac{GM}{r} - \frac{\varpi\varOmega_{\ast}B_rB_\phi}{4\upi\rho v_r}
            \end{equation}
            is conserved, i.e., $\partial\mathcal{B}_{\rm eq}/\partial r = 0$ \citep{Lamers1999}.  The first term in this quantity is the kinetic energy flux, the second is the thermal energy flux, the third is gravitational, and the fourth is magnetic.  We note that the thermal term is negative (because $\rho$ decreases with radius) and the fourth is positive (because $B_\phi<0$).
            
            We generalize this conservation law to the non-equatorial case by defining the (conserved) energy flux across a closed spherical surface to be:
            \begin{equation}
                \dot{E} = \oint r^2\rho v_r \mathcal{B}_{\rm eq}\:\mathrm{d}\Omega.
            \end{equation}
            Formally, $\dot{E}$ is only constant for an equatorial rotator, however, we find that the deviation from constant $\dot{E}$ is small.    We plot the energy loss rate in panel (c) of Figure~\ref{fig:heatmap}.% (see Figure~\ref{fig:constants}).
            
            Analytically (e.g., \citealt{Metzger2007}), we expect the energy loss rate in the fast magnetic rotator (FMR) regime (where magnetocentrifugal forces dominate; \citealt{Lamers1999}) to scale as
            \begin{equation}
                \dot{E}_{\rm FMR} \sim \frac{3}{2}\dot{M}\eta v_{\rm M}^2,\label{eq:edot_FMR}
            \end{equation}
            where $\eta=\varOmega\dot{J}/\dot{E}$ is the ratio of spindown power to asymptotic wind power, and
            \begin{equation}
                v_{\rm M} = \left(\frac{\varPhi_{\rm open}^2\Omega_\ast^2}{\dot{M}c^3}\right)^{1/3}c = \sigma^{1/3}c
            \end{equation}
            is the Michel velocity \citep{Michel1969}, $\sigma$ is the wind magnetization \citep{Lamers1999,Metzger2007}, and $\varPhi_{\rm open}$ is the open magnetic field flux (defined below).  For most FMR cases, $\eta\approx1$, but for FMRs with small $B_{\ast}$ and and short periods, most of the spindown power is used to unbind the wind, and thus $\eta\gg 1$ \citep{Metzger2011}.  For thermally driven cases (where magnetocentrifugal forces are negligible), $\eta\ll1$, because the spindown power does not make an appreciable contribution to the asymptotic wind power.  Because our parameter space spans the regime from thermally driven winds to FMRs, this thermal component makes a large contribution to the total asymptotic wind power in many of our simulations.  We find that the addition of an additional thermal term,
            \begin{equation}
                \dot{E}_{\rm th} \simeq \dot{M}c_T^2,
            \end{equation}
            provides a more accurate estimate to our measured energy fluxes.
            
            The wind magnetization $\sigma$ is also a measure of how relativistic the wind, as $\sigma\sim S$, where $S$ is the Poynting flux.  We typically say the wind is relativistic when $\sigma>1$.  We do find that a small number of our simulations (those with the smallest $\xi_T$ and and largest $\xi_\varOmega$) have $\sigma>1$, however, the effects of relativity on these winds is outside the scope of this paper.
            
            \subsubsection{Alfv\'en Radius}
            In a 1D magnetized wind, the Alfv\'en point is the point where the wind velocity is equal to the Alfv\'en velocity, given by
            \begin{equation}
                v_A = \frac{B^2}{4\pi\rho}.
            \end{equation}
            In higher dimensions, we define an an Alfv\'en \emph{surface} $\mathcal{S}_A$ that is the locus of all points where $v_p = v_A$, where
            \begin{equation}
                v_p^2 = v_r^2 + v_\theta^2 \label{eq:vpoloidal}
            \end{equation}
            is the poloidal velocity.  We also define $B$ using the poloidal field strength $B_p$, defined analogously to the poloidal velocity.  In principle, we could numerically solve for $\mathcal{S}_A$ in our simulation data and then use that to compute an average Alfv\'en radius $\langle R_A\rangle$.  However, it is far more convenient to determine $\langle R_A\rangle$ from the eigenvalues of the problem i.e., $\dot{M}$ and $\dot{J}$.  Consider that the angular momentum loss rate can also be written as
            \begin{equation}
                \dot{J} = \dot{M}R_A^2\Omega_\ast.
            \end{equation}
            Since we have already computed $\dot{M}$ and $\dot{J}$ in the previous sections, it is straightforward to use them to compute an average Alfv\'en radius,
            \begin{equation}
                \langle R_A\rangle = \sqrt{\frac{\dot{J}}{\dot{M}\Omega_\ast}}.\label{eq:meanra}
            \end{equation}
            We plot this quantity in panel (d) of Figure~\ref{fig:heatmap}.
            
            \subsubsection{Open Magnetic Field Flux}
            The unsigned open magnetic field flux is given by:
            \begin{equation}
                \varPhi_{\rm open} = \oint r^2 |B_r|\:\mathrm{d}\Omega.
            \end{equation}
            At large radii, this quantity is constant, because all field lines are open.  However, close to the surface, a region of closed magnetic field lines can develop, reducing $\varPhi_{\rm open}$.  The ratio of $\varPhi_{\rm open}$ at these two locations, then, provides a measure of the shape of the magnetic field.  As $\varPhi_{\rm open}/\varPhi_{\rm open,0}$ increases, the surface magnetic field becomes more and more open, with $\varPhi_{\rm open}/\varPhi_{\rm open,0}=1$ corresponding to a split-monopole field.  We plot this ratio in panel (f) of Figure~\ref{fig:heatmap}.
        
        \subsection{Structure and Evolution of the Magnetized Winds}\label{sec:structure}
\begin{figure*}
                \centering
                \includegraphics[width=\textwidth]{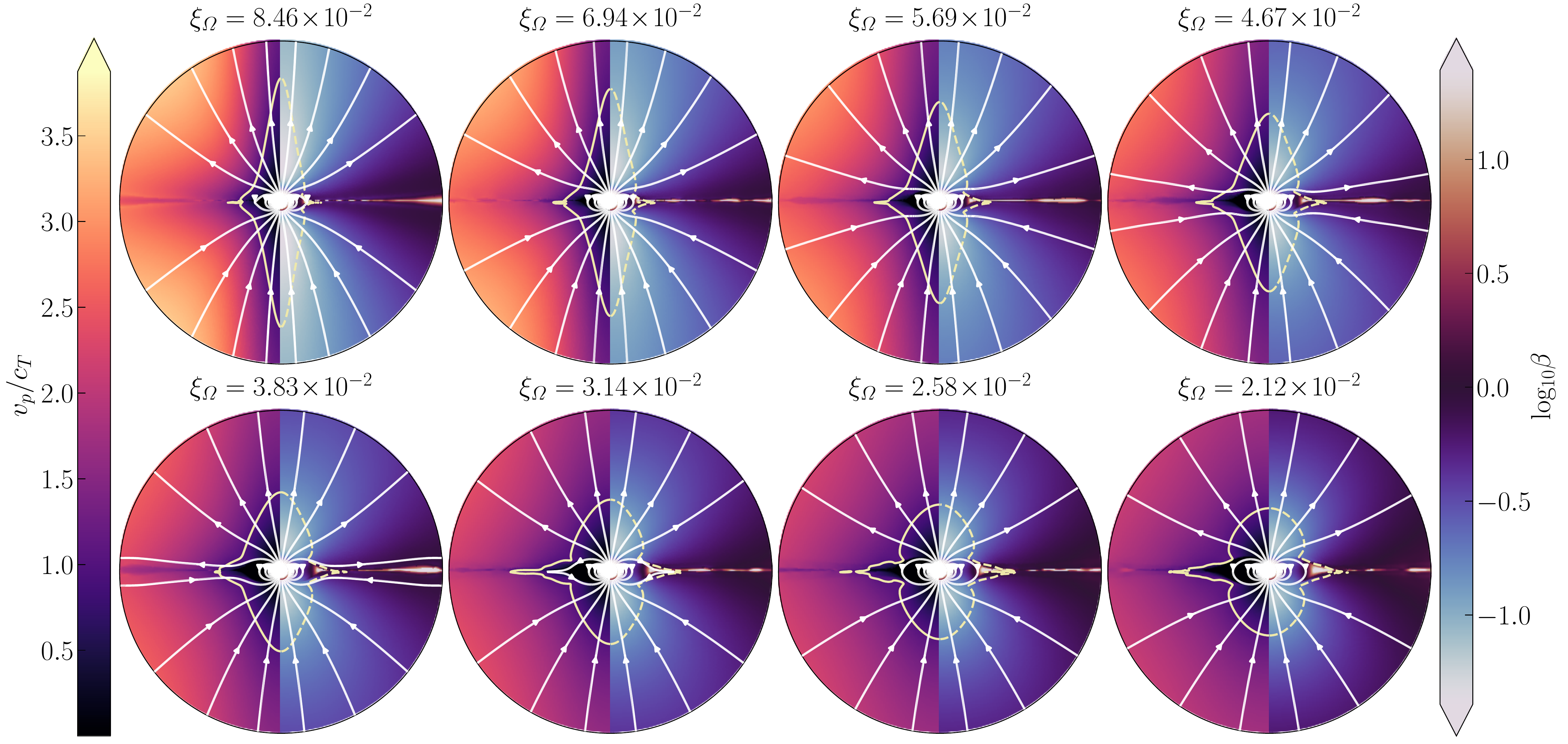}
                \caption{The inner $20\rast$ of 2D axisymmetric simulations at various rotation periods for a fixed sound speed ${\xi_T=0.1641}$.  On the left hand side the poloidal mach number $\mathcal{M}=v_{p}/c_{T}$ is shown; on the right hand side the plasma $\beta$ is shown.  The contours in yellow show the sonic surfaces: the mach surface $\surf_{\mathcal{M}}$ on the LHS, and the 
                %~ Alfv\'en surface  $\surf_{A}$ (solid) and the fast and 
                slow magnetosonic surface $\surf_{-}$ (dashed) on the RHS.  The magnetic field lines are shown in white on both sides.  As the rotation period is lowered, the centrifugal forces become larger and the sonic surfaces become more cylindrical.  At the same time, the region of closed magnetic field lines becomes bigger.}
                \label{fig:vsP_lowcT}
            \end{figure*}

\begin{figure*}
                \centering
                \includegraphics[width=\textwidth]{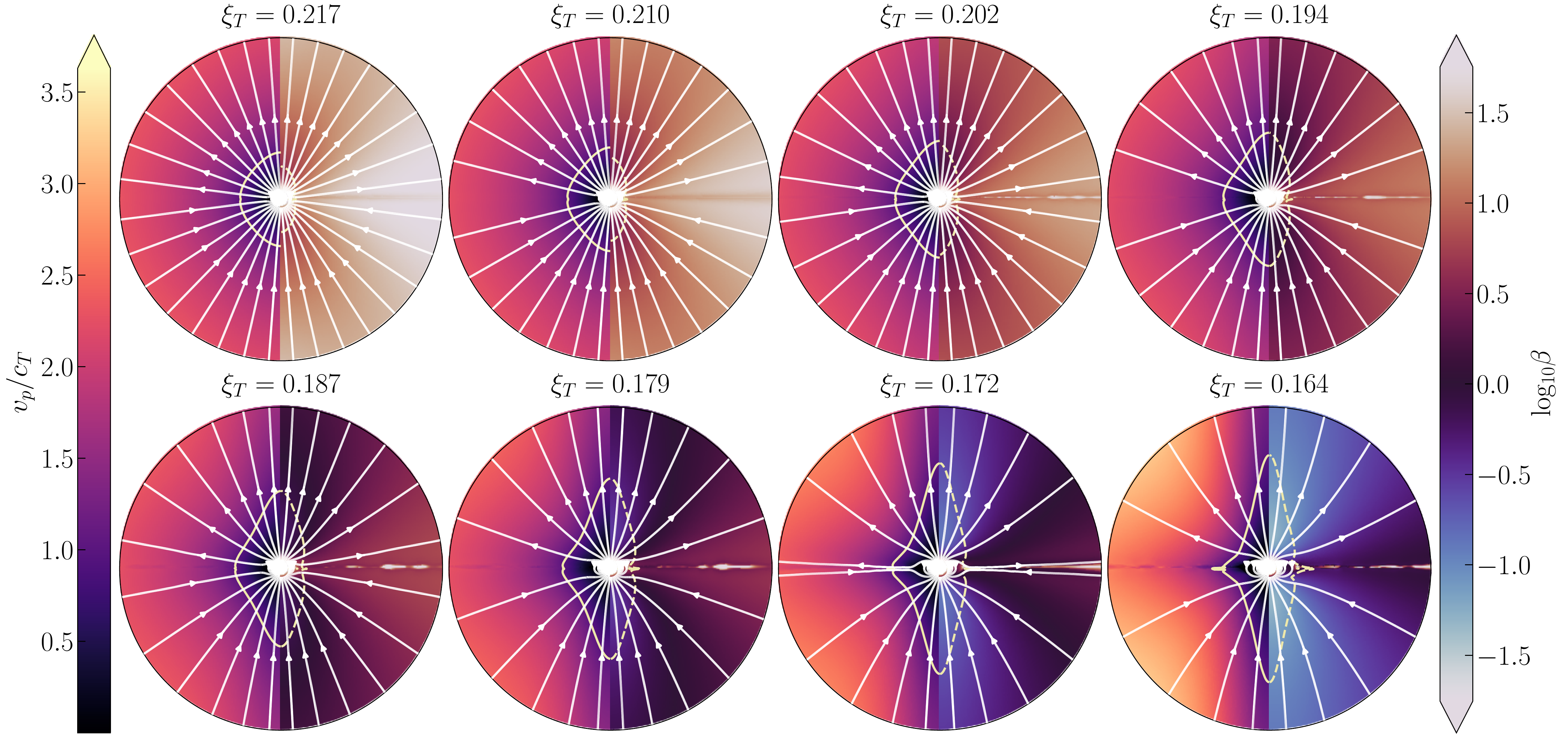}
                \caption{The same as Figure~\ref{fig:vsP_lowcT}, but for varying sound speed at a fixed, moderate rotation rate $\xi_{\varOmega}=\num{6.94e-2}$.  As the sound speed is lowered, the simulation becomes more magnetically dominated and sonic surfaces move to larger radii.  The mach surface becomes more oblate, and a pronounced cusp appears in all surfaces on the equator.  Furthermore, we note that as the sound speed decreases, the wind is more and more directed into two off-equatorial lobes, and a region of closed magnetic field lines forms in the subsonic region around the equator.}
                \label{fig:vscT_medP}
            \end{figure*}

\begin{figure*}
                \centering{}
                \includegraphics[width=\textwidth]{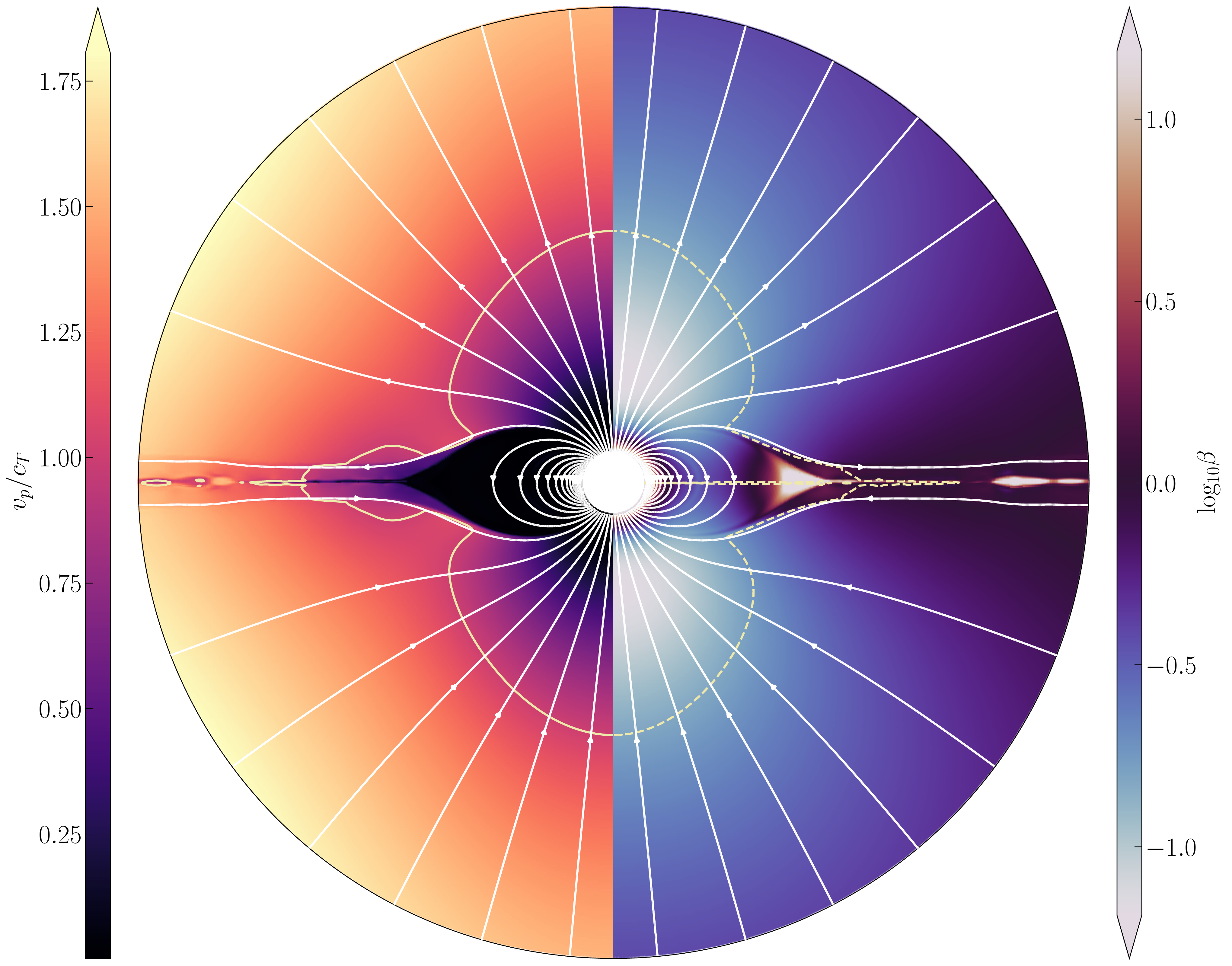}
                \caption{Zoom-in to the inner $15\rast$ of the 2D simulation with $\xi_T=0.1641$ and $\xi_{\varOmega}=\num{2.12e-2}$.  From this plot, it is clear that the poloidal velocity of the wind in the closed zone is zero (or nearly zero).  Furthermore, as the closed field lines come to a cusp at the edge of the closed zone, the plasma $\beta$ increases sharply.}
                \label{fig:zoom}
            \end{figure*}

            Here, we examine a selection of our 2D simulations in more detail.
            %~ The final snapshots of these simulations are presented 
            In Figures~\ref{fig:vsP_lowcT} and \ref{fig:vscT_medP}, we show the poloidal mach number $v_p/c_T$ and plasma $\beta$, defined as the ratio of thermal pressure to magnetic pressure:
            \begin{equation}
                \beta = \frac{2c_T^2\rho}{B^2},
            \end{equation}
            in the final snapshots of our simulations.
            %The parameters of our full list of simulations is given in Appendix~\ref{app:fullsimtable}, and additional snapshots of our simulations are presented in Appendix~\ref{app:Figures}.  
            We plot the inner $20\rast\times20\rast$ (40 percent of the total radial extent) region for each model in each of these figures, which is enough to capture the sonic surfaces (see below) at the equator in all cases, though the sonic surfaces at the poles is sometimes outside this region.  We also present a zoomed in $(10\rast\times10\rast)$ panel of a single one of the simulations in Figure~\ref{fig:zoom}, to better show the structure of the magnetic field near the surface of the star.
            
            % \begin{table*}
            %     \centering
            %     \input{array-reduced.tex}
            %     \caption{The dimensionless isothermal sound speed, core rotation rate, magnetic field strength, mass loss rate, angular momentum loss rate, energy loss rate, spindown time, and open magnetic flux for the 2D simulations discussed in this section.% (the table of parameters for the full list of 2D simulations can be found in Appendix~\ref{app:fullsimtable})
            %     The first half of the table shows the simulations used for plots where $P_{\ast}$ is varied, and the second half shows the simulations used for plots where $\xi_T$ is varied (repeated simulations are not listed twice).  All quantities are scaled to the proto-magnetar wind regime (see Table~\ref{tab:scaling} and Appendix~\ref{app:units} for more details}
            %     \label{tab:2d-sims}
            % \end{table*}
            
            We define four \emph{sonic surfaces} in these plots.  On the Mach surface $\surf_\mathcal{M}$, the poloidal speed of the wind $v_p$ is equal to the isothermal sound speed $c_T$.  On the Alfv\'en surface $\surf_{A}$, the poloidal speed is equal to the Alfv\'en speed $v_A$.  On the fast and slow magnetosonic surfaces $\surf_{\pm}$, the poloidal speed is equal to the fast or slow magnetosonic speed:
            \begin{equation}
                v_{p}^{2} = v^{2}_{\pm} \equiv \frac{1}{2}\left(v_{A}^{2}+c_{T}^{2}\pm\sqrt{(v_{A}^{2}+c_{T}^{2})^{2}-4v_{A}^{2}c_{T}^{2}\cos^{2}\vartheta}\right),\label{eq:surfms}
            \end{equation}
            where $\vartheta$ is the angle between the magnetic field and the direction of wave propagation.  For the purposes of determining $\mathcal{S}_{\pm}$, we assume that magnetosonic waves propagate radially, i.e.,
            \begin{equation}
                \cos\vartheta = \frac{\mathbfit{B}\cdot\uvec{r}}{B}.
            \end{equation}
            The magnetic forces are more dynamically important at low sound speeds than they are at higher sound speeds, where thermal effects dominate.  Thus, at lower $\xi_T$, the sonic surfaces $\surf_{A},\surf_{\pm}$ move outwards.  The Mach surface $\surf_{\mathcal{M}}$ also naturally moves outward at lower sound speed as $r_{s}\sim c_{T}^{-2}$. At larger $\xi_{\varOmega}$, the rotational velocity, which decreases with the \emph{cylindrical} radius $\varpi$, becomes more dynamically important.  Thus, the sonic surfaces take on a more cylindrical quality, moving further out at the poles than at the equator \citep{Keppens2000}.
            
\begin{figure*}
                \centering
                \includegraphics[width=0.5\textwidth]{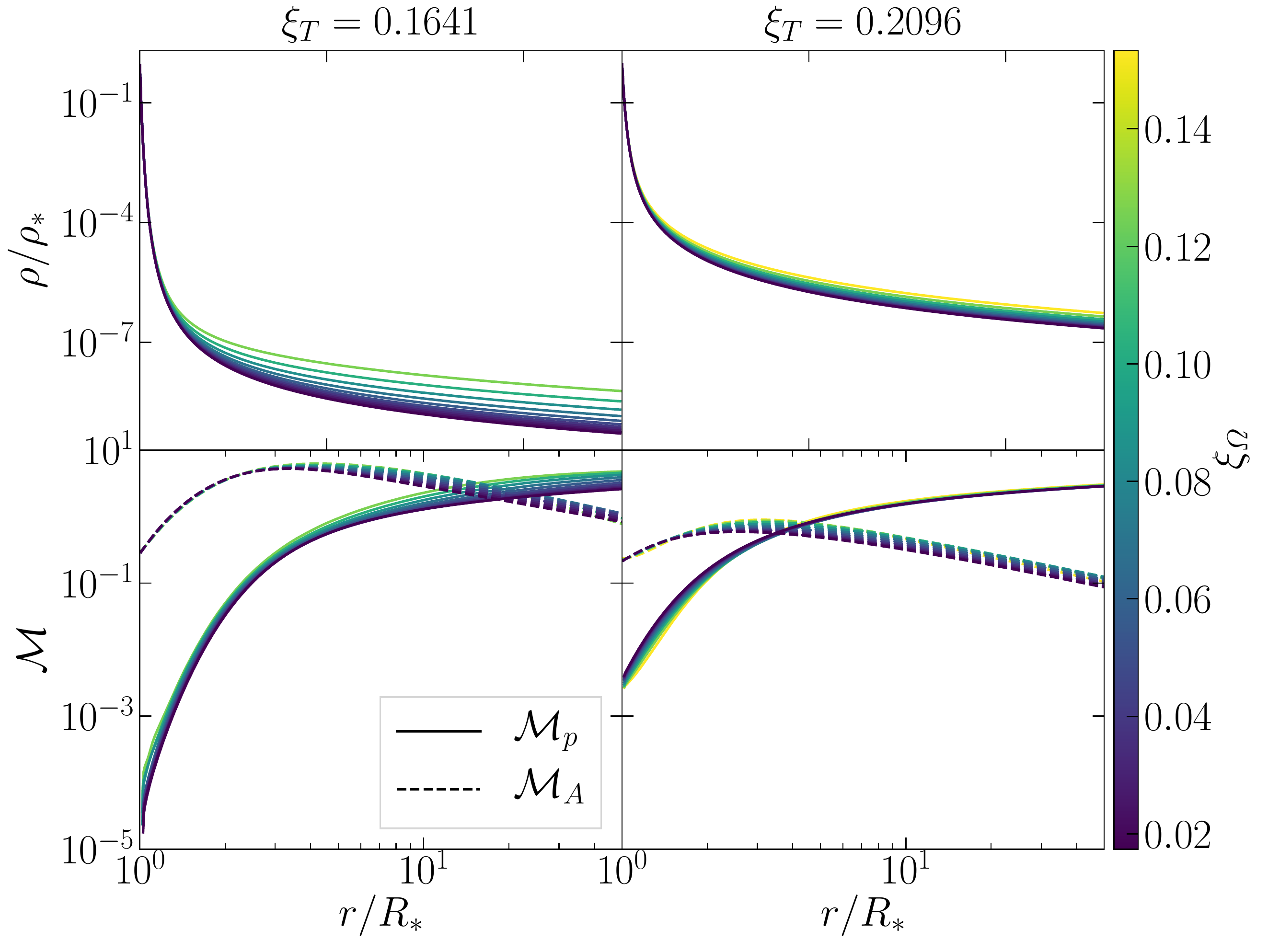}~
                \includegraphics[width=0.5\textwidth]{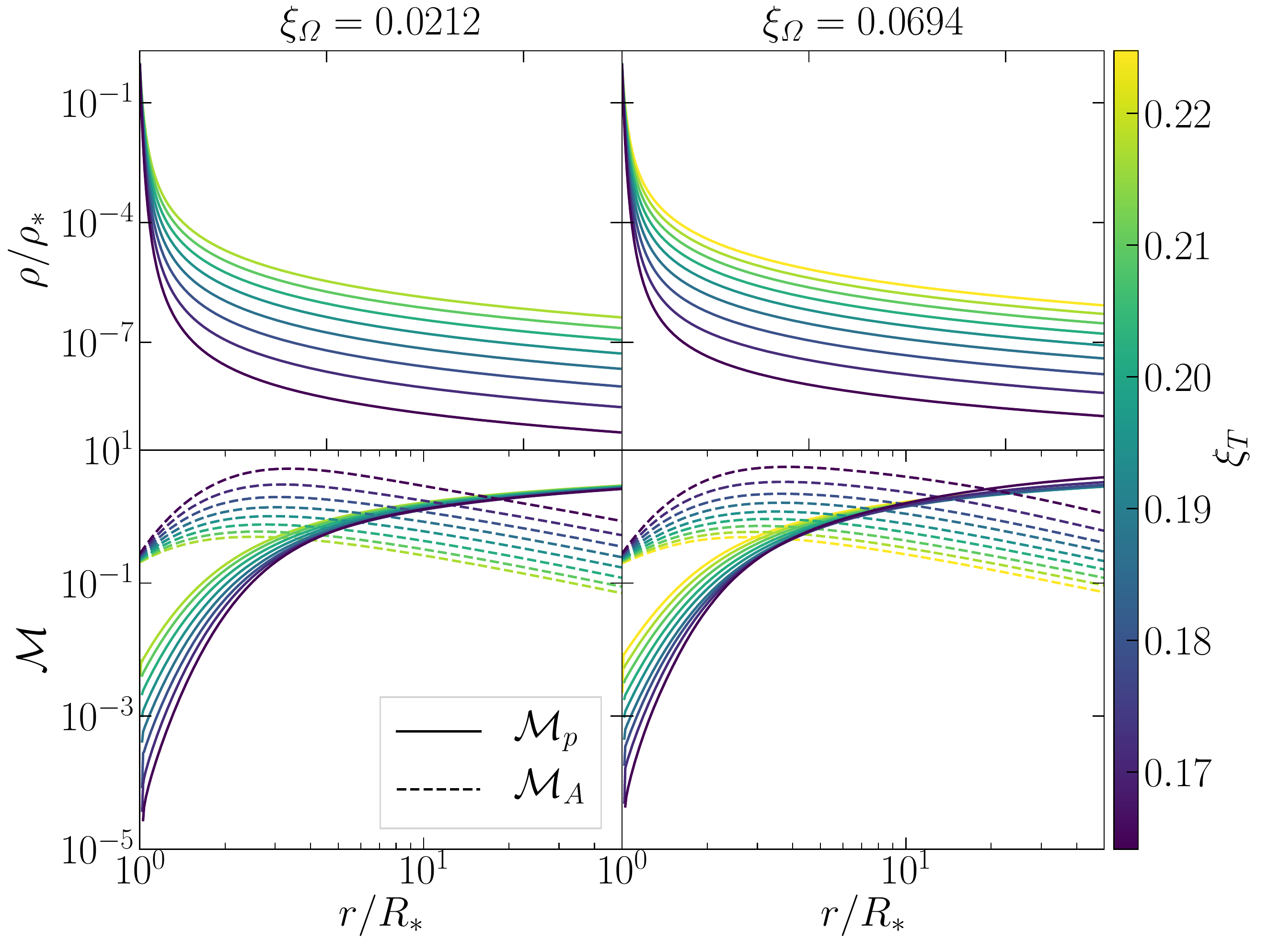}
                \caption{Density and mach number ($\mathcal{M}=v/c_T$ profiles for simulations at varying $\xi_T$ and $\xi_\varOmega$.  The left two panels show constant $\xi_T$ and varying $\xi_\varOmega$, while the right two panel show the reverse.  Profiles are taken at $\theta=\pi/8$.  For cooler models, rotation has a larger impact on the structure of the wind.  In particular, more rapidly rotating models have larger $\rho$ and $\mathcal{M}_p$, with both effects diminishing at larger $\xi_T$.}
                \label{fig:profiles}
            \end{figure*}

            In Figure~\ref{fig:profiles}, we plot the density and Mach number profiles of a subset of our simulations, where the Mach number is defined as
            \begin{equation}
                \mathcal{M} = \frac{v}{c_T},
            \end{equation}
            for some characteristic velocity $v$ (e.g., $\mathcal{M}_p = v_p/c_T$ is the poloidal Mach number and $\mathcal{M}_A = v_A/c_T$ is the Alfv\'enic Mach number).  In each set of panels, each column corresponds to the simulations shown in Figures~\ref{fig:vsP_lowcT} and \ref{fig:vscT_medP}, respectively. We see the density profile is highly dependent on both the sound speed and the rotation rate, with cooler (i.e., lower $\xi_T$) models seeing a relatively larger effect from rotation.  Hot models also have larger outflow velocities than cooler models do.  Rapid rotation also changes the shape of the Mach number profile -- as the rate of rotation increases, $\mathcal{M}_p$ increases at small radii, but remains relatively constant at large radii.
        
            % \begin{figure*}
            %     \centering
            %     \includegraphics[width=\textwidth]{plots/x1a.pdf}
            %     \caption{The same as Figure~\ref{fig:vsP_lowcT}, except with a larger $\xi_T = 0.1793$.  In this simulation, the gas pressure is larger, so the relative contribution of the magnetic field is lessened.  As such, the sonic surfaces are more spherical, and  $S_{A},S_{+}$ are located at smaller radii, as the period increases.}
            %     \label{fig:vsP_medcT}
            % \end{figure*}
            
            % \begin{figure*}
            %     \centering
            %     \includegraphics[width=\textwidth]{plots/f1x.pdf}
            %     \caption{The same as Figure~\ref{fig:vsP_lowcT}, but for varying sound speed at a fixed, rotation rate $\xi_{\varOmega} = \num{1.74e-2}$.  As the sound speed is lowered, the simulation becomes more magnetically dominated and sonic surfaces move to larger radii.  The mach surface becomes more oblate, and a pronounced cusp appears in all surfaces on the equator.  Furthermore, in the lowest sound speed cases, a pronounced area of closed magnetic field lines forms in the subsonic region around the equator.}
            %     \label{fig:vscT_longP}
            % \end{figure*}
            
        \subsection{Effect of Magnetosphere Tilt}
        \label{sec:tilt}
\begin{figure*}
                \centering
                \includegraphics[width=\textwidth]{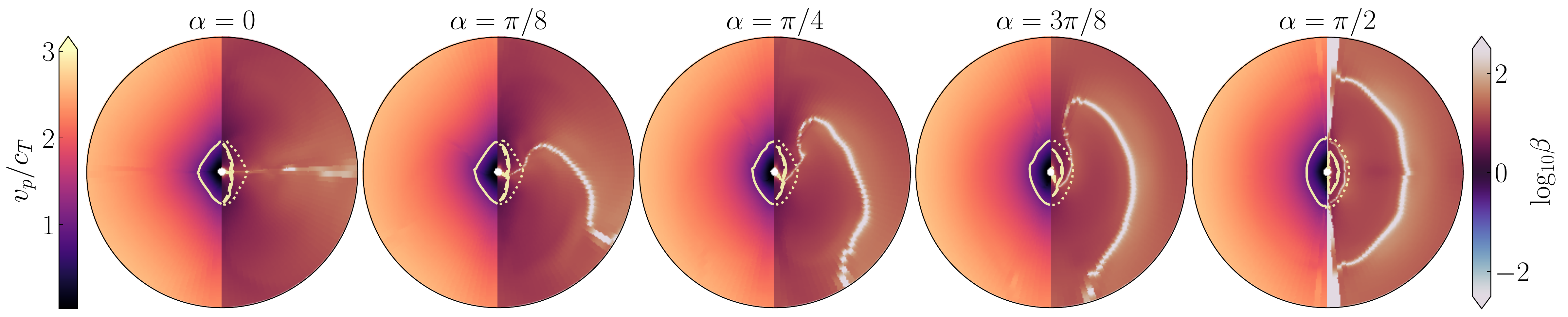}
                \caption{The inner $30\rast$ of the $\phi'=0$ slice of 3D simulations with varying magnetosphere tilt $\alpha$ at fixed rotation rate $\xi_{\varOmega} = \num{0.085}$ and sound speed $\xi_T = 0.210$.  Contours depict the sonic surfaces as described in Figure~\ref{fig:vsP_lowcT}, with the addition of the Alfv\'en surface (solid) and fast magnetosonic surface (dotted).  We note that even in the aligned $(\alpha=0)$ simulation, a small tilt develops due to spontaneous symmetry breaking.  The magnitude of this effect, and the resultant deviation from 2D simulations, is greater for cooler and more rapidly rotating simulations (see Figure~\ref{fig:vsalpha_params}); for the simulation in question it leads to a $\sim$0.5 per-cent deviation in $\dot{M}$ and $\dot{E}$ and a $\sim$10 per-cent deviation in $\dot{J}$ and $\varPhi$.}
                \label{fig:vsalpha_716}
            \end{figure*}

            We simulated a smaller number of 3D models with tilt angles $\alpha>0$, at representative regions of parameter space.  Snapshots of some of these simulations are presented in Figure~\ref{fig:vsalpha_716}.  We note that the $\alpha=0$ case is not identical to the 2D simulation of the same $\xi_T$ and $\xi_\varOmega$.  There is a small asymmetry in the current sheet, which, in our 2D simulations lies exactly along the equator.  The degree of asymmetry increases for cooler and more rapidly rotating (smaller $\xi_T$ and larger $\xi_\varOmega$) simulations, and leads to significant uncertainties in $\dot{M}$, $\dot{J}$, and $\dot{E}$.
        
            \begin{figure*}
                \centering
                \includegraphics[width=\textwidth]{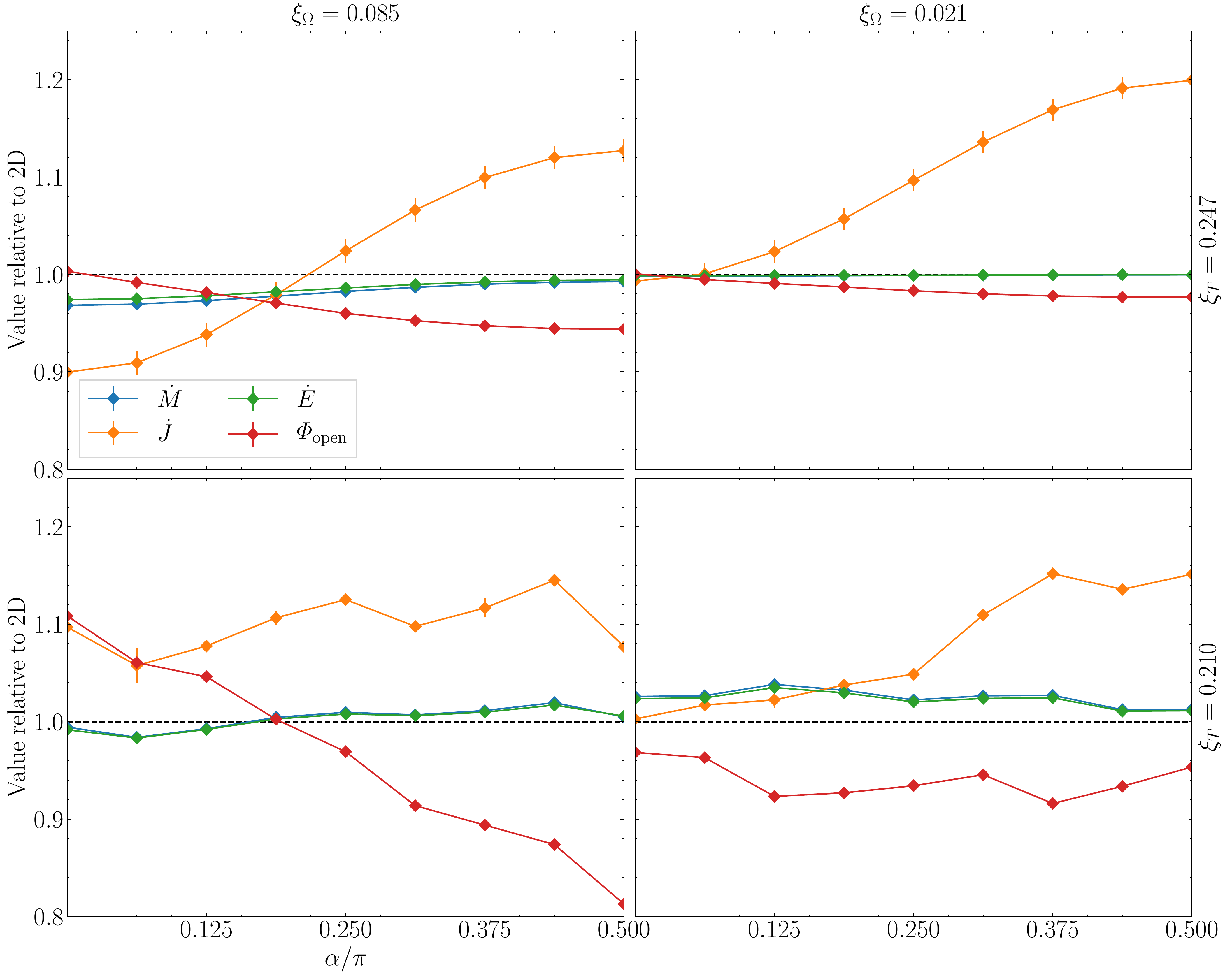}
                \caption{$\dot{M}$, $\dot{J}$, $\dot{E}$, and $\varPhi_{\rm open}$ (normalized by their value in the aligned case) as a function of the tilt angle $\alpha$ for a series of 3D simulations at different $\xi_T$ and $\xi_\varOmega$.  We identify broad trends in these results -- $\dot{J}$ tends to decrease as $\alpha$ increases, while $\varPhi$ decreases.  $\dot{M}$ and $\dot{J}$, on the other hand, stay relatively constant.  These trends break down in cooler and more rapidly rotating simulations.}
                \label{fig:vsalpha_params}
            \end{figure*}
            
            In Figure~\ref{fig:vsalpha_params}, we investigate how $\dot{M}$, $\dot{J}$, $\dot{E}$, and $\varPhi_{\rm open}$ are affected by the tilt, in simulations with 4 different combinations of $\xi_T$ and $\xi_\varOmega$.  We find that $\dot{M}$, and $\dot{E}$ are very nearly constant with tilt angle (at most, varying by $\sim$1 per-cent over the full range of $\alpha$).  By contrast, $\dot{J}$ and $\varPhi_{\rm open}$ show strong trends with $\alpha$.  This implies that the trends in these quantities are driven by intrinsic effects of the misaligned magnetic field, rather than as a consequence of a trend in the mass loss rate.  However, these trends break down in our low $\xi_T$ simulations, shown in the lower two panels of the figure.

             In the simplest model of the magnetic field, where the field lines are purely radial, we would expect $\varPhi_{\rm open}$ to remain constant with $\alpha$, as the magnetic field strength is independent of angle.  For higher order moments, the polar field strength is twice the equatorial field strength, and $\varPhi_{\rm open}$ starts to vary with $\alpha$ as different parts of the magnetosphere are opened by the wind and by centrifugal forces.  For the aligned $(\alpha=0)$ case, the open field lines are located near the poles, where $\dot{J}=0$.  As the tilt angle increases, more open field lines are equatorial, and $\dot{J}$ along open field lines increases.  As mass is ejected only along open field lines, we would then expect that larger tilt angles would mean larger $\dot{J}$. As for $\dot{E}$, we would expect that for more thermally dominated simulations, such as the ones shown in Figure~\ref{fig:vsalpha_params}, the trends in $\dot{E}$ will closely follow the trends in $\dot{M}$ (as $\dot{E}\sim\dot{M}c_T^2$ in thermally dominated winds).  In the FMR regime, we would expect the trend on $\dot{E}$ to track the trends in $\varPhi_{\rm open}$.  However, a full study of how these quantities vary over the complete three-dimensional parameter space of $(\xi_T,\xi_\varOmega,\alpha)$ is beyond the scope of this paper.
        
        \subsection{Resolution Effects}
        \label{sec:results_resolution}
\begin{figure}
                \centering
                \includegraphics[width=\linewidth]{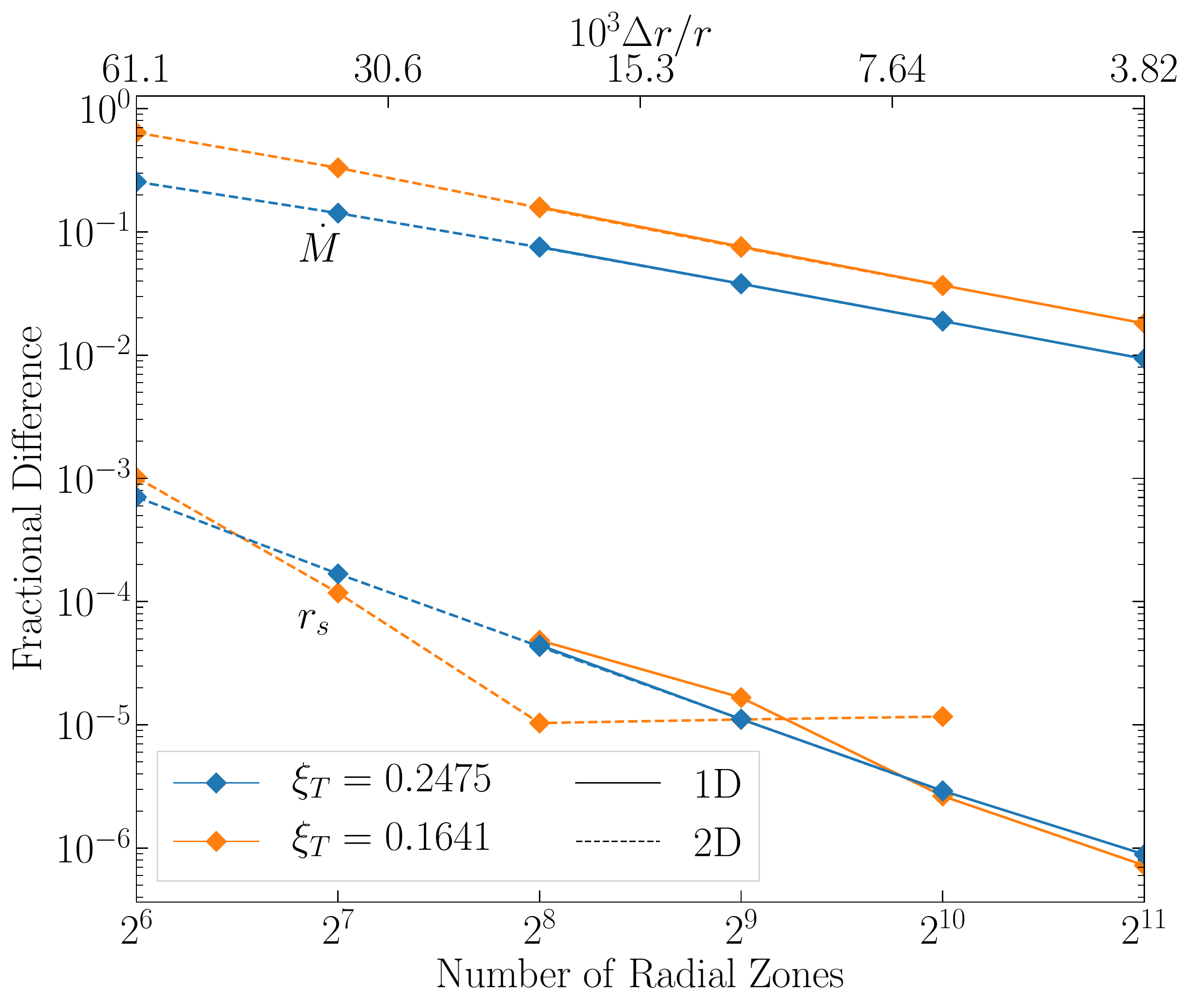}
                \caption{Fractional difference in $\dot{M}$ and $r_s$ (compared to the analytic values as given by Equations~\eqref{eq:rsonic} and \eqref{eq:mdot-an}) vs the number of radial zones (or, equivalently, the fractional spacing $\Delta r/r$, shown at top), for ``high'' ($\xi_T=0.2475$) and ``low'' ($\xi_T=0.1641$) isothermal sound speeds, for the 1D (solid) and 2D (dashed) NRNM simulations.  The mass accretion rate converges much more slowly than the sonic radius, and to lower accuracy.  However, the sound speed does not meaningfully affect the rate of convergence.}
                \label{fig:res}
            \end{figure}

            In Figure~\ref{fig:res}, we examine the effects of radial resolution on our measurement of $\dot{M}$ and the sonic radius $r_s$ in our NRNM simulations, for 1D and 2D simulations.  We find that our measurements of the sonic radius converge more quickly with increasing radial resolution than our measurements of the mass loss rate.  Furthermore, the move from one to two spatial dimensions does not significantly affect the rate of convergence\footnote{However, in 2D, and especially in 3D, we are forced to run at lower resolution in order to maintain reasonable simulation walltimes, which limits the resolution that is practically available to us.}.
            
\begin{figure}
                \centering
                \includegraphics[width=\linewidth]{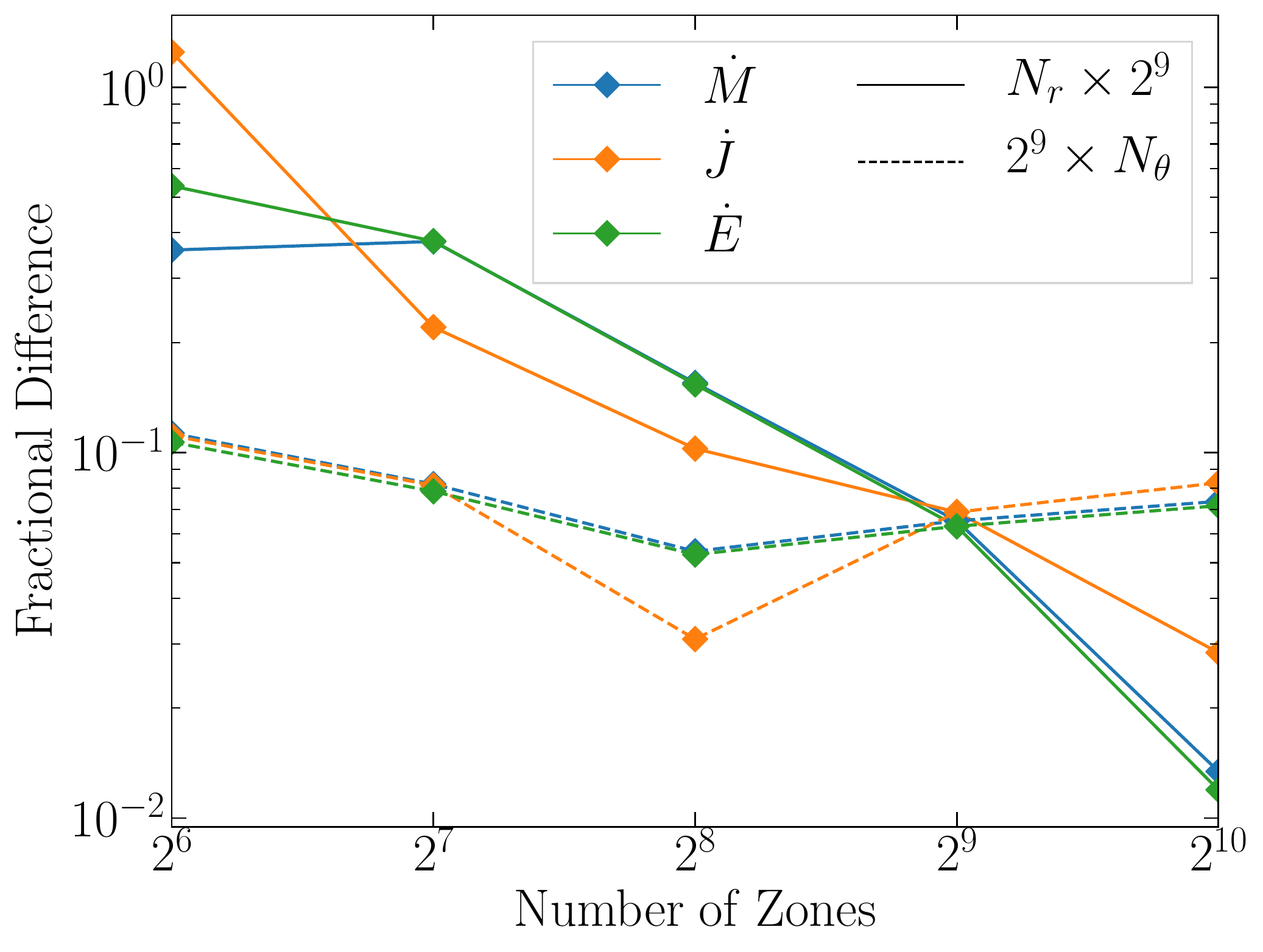}
                \caption{Fractional difference in $\dot{M}$, $\dot{J}$, and $\dot{E}$, and $r_s$ vs the number of zones for 2D simulations with $\xi_T=0.1641,$ $\xi_{\varOmega}=\num{2.12e-2},$ and $\xi_{B}=\num{4.63e-2}$.  Solid lines show the case of fixed angular resolution (at the fiducial resolution specified in \S\ref{sec:Methods_Resolution}) and varying radial resolution, while the dashed lines show the case of fixed radial resolution and varying angular resolution.  We see that increasing the angular resolution has only a marginal impact on the precision of our measurements; simulating with high radial resolution is of far greater importance in making precise measurements of the wind properties.}
                \label{fig:res-rm}
            \end{figure}

            In Figure~\ref{fig:res-rm}, we examine the convergence of 2D simulations with $\xi_T=0.1641,$ $\xi_{\varOmega}=\num{2.12e-2},$ and $\xi_{B}=\num{4.63e-2}$ (the same parameters as the simulation shown in Figure~\ref{fig:zoom}).  We find that the wind properties $\dot{M},$ $\dot{J}$, and $\dot{E}$ readily converge with increasing radial resolution, but stay relatively constant with increasing angular resolution, consistent with the NRNM expectations.  This result is consistent for both high and low $\xi_T$ and $\xi_\varOmega$.  This helps justify our use of relatively lower angular resolution in our 3D models.
            
    \section{Discussion}\label{sec:Discussion}
    
    The results shown in the previous section can be scaled to a number of physical systems of interest, by using different values of $\rho_{\ast}$, $G\mast$, and $\rast$ when converting from code units to physical units.  This is discussed in more detail in Appendix~\ref{app:units}.  Because the field strength and rotation rate are specified with the dimensionless parameters $\xi_B$ (Equation~\ref{eq:mag}) and $\xi_{\varOmega}$ (Equation~\ref{eq:rot}), they also naturally scale with our choice of units (though not necessarily to values appropriate for the regime of interest.  We summarize the scalings we use in Table~\ref{tab:scaling}.
    
    \begin{table*}
        \centering
        \begin{tabular}{c c c c c c}
            \hline
            Quantity & Code Units & Proto-Magnetar & Sun-like Star$^{\rm a}$ & Hot Jupiter$^{\rm b}$ \\\hline
            $\rho_{\ast}$ & 1 & \SI{e11}{\gram\per\cm\cubed} & \SI{e-16}{\gram\per\cm\cubed} & \SI{e-15}{\gram\per\cm\cubed}\tstrut\\
            $\mast^{\rm c}$ & $1/G$ & \SI{1.4}{\msun} & \SI{1}{\msun} &  \SI{0.7}{M_J}\\
            $\rast$ & 1 & \SI{10}{\km} & \SI{1}{R_\odot} & \SI{1.4}{R_J}\bstrut\\\hline
            $B_{\ast}$ & 0.0655 & \SI{e15}{\gauss} & \SI{0.101}{\gauss}  & \SI{0.0221}{\gauss}\tstrut\\
            $P_{\ast}$ & \numrange{16}{256} & \numrange{1.2}{19}\,\si{\ms} & \numrange{0.29}{4.7}\,\si{\day} & \numrange{14.5}{231}\,\si{\hour}\\
            $c_T$ &
            \numrange{0.2}{0.35} & \numrange{0.091}{0.16}\,$c$ & \numrange{87}{150}\,\si{\km\per\second} & \numrange{6.02}{10.5}\,\si{\km\per\second}\\
            \hline
        \end{tabular}
        \caption{Summary of the scaling factors to the various contexts considered here.  In addition to $\rho_{\ast}$, $GM$, and $R$, which together define the unit conversion, we also show the range of values for $B_{\ast}$, $P_{\ast}$, and $c_T$, as determined by the $\xi_B$, $\xi_\varOmega$, and $\xi_T$ ranges given in \S\ref{sec:propwinds}.  Since the ranges of these values were deliberately chosen to cover the proto-magnetar regime, they do not always produce appropriate values in other contexts.\newline{}
        $^{\rm a}$ $\rho_{\ast}$ in this case is chosen such that $\dot{M}\sim\dot{M}_{\odot}\simeq\SI{2e12}{\g\per\s}$ for a $\xi_{B}$ and $\xi_{\varOmega}$ similar to that used in \citet{Finley2017}.  See \S\ref{sec:solar} for details.\newline
        $^{\rm b}$ Using the \citet{MurrayClay2009} values for $\rho_\ast$, $\mast$, and $\rast$\newline
        $^{\rm c}$ The quantity $G\mast$ is specified in the code; $\mast$ is presented here on its own for the sake of readability.}
        \label{tab:scaling}
    \end{table*}
    
    Below, we focus on the application to newly-born highly-magnetic and rapidly-rotating proto-neutron stars, and then discuss other applications of our results.
    
    %\FloatBarrier
    \subsection{Analytic Approximations}\label{sec:analytics}
        For the purpose of analytical estimation, we closely follow the analysis of \citet{Thompson2003}.  Consider a star with a simplified magnetic field structure with
        \begin{equation}
            B_{p} = B_{\ast}\left(\frac{\rast}{r}\right)^{\lambda},
        \end{equation}
        For a split monopole field structure, (e.g., \citealt{Weber1967}), $\lambda=2$, and for a dipole field, $\lambda=3$.  Therefore, we should expect that some $2<\lambda<3$ to best describe the magnetic field structure of our simulations, depending on the sound speed $c_T$. From this equation, we find that
        \begin{equation}
            R_{A}^{2\lambda-2} = B_{\ast}^{2}\rast^{2\lambda}\dot{M}^{-1}v_{A}^{-1},
        \end{equation}
        assuming spherical outflow (i.e., $\dot{M}=4\upi\rho v_{r}$ and $v_{p}=v_{r}$).  
        
        As discussed in \cite{Lamers1999,Thompson2004,Metzger2007},  basic scalings for magnetocentrifugal spindown is provided by assuming effective co-rotation of the wind material out to the Alfv\'en point, where 
        \begin{equation}
            R_A^2\varOmega_{\ast}^2 = \frac{3}{2}\eta v_{\rm M}^2 = \frac{3}{2}\eta\left(\frac{\rast^4B_\ast^2\Omega^2}{\dot{M}}\right)^2,
        \end{equation}
        For a spherical wind, the Alfv\'en speed can thus be estimated as
        \begin{equation}
            v_{A} = \frac{v_{\rm M}^{3}}{R_{A}^{2}\varOmega^{2}_{\ast}} = \left(\frac{2}{3\eta}\right)^{3/2}R_{A}\varOmega_{\ast}.
        \end{equation}
        This can be incorporated into our estimate for the Alfv\'en radius,
        \begin{equation}
            R_{A}^{2\lambda-1} = \left(\frac{3\eta}{2}\right)^{3/2}B_{\ast}^{2}\rast^{2\lambda}\dot{M}^{-1}\varOmega_{\ast}^{-1}
        \end{equation}
        
        We can use this framework to estimate the angular momentum loss rate, and thus, the spindown time.  Starting with Equation~\eqref{eq:jdot}, and assuming that $v_{\phi}=R_{A}\varOmega_{\ast}$ and $\left.{B_{\phi}}/{B_{r}}\right|_{\surf_{A}}\ll 1$, we find
        \begin{equation}
            \dot{J} \approx \frac{\upi}{4}\dot{M}R_{A}^{2}\varOmega_{\ast}.\label{eq:jdot-an}
        \end{equation}
        Once again assuming the star is a uniform density sphere, we find the spindown time to be
        \begin{equation}
            \tau_{J} = \frac{8}{5\upi}\mast\left[\dot{M}^{3-2\lambda}B_{\ast}^{-4}\rast^{-2}\varOmega_{\ast}^{2}\left(\frac{3\eta}{2}\right)^{-3}\right]^{1/(2\lambda-1)}.\label{eq:tauJ-an}
        \end{equation}
        Scaling these to values appropriate to the proto-magnetar regime (see Table~\ref{tab:scaling}), we find
        \begin{equation}
            \tau_{J} \approx
            \begin{cases}
                30 \eta^{-1} B_{\ast,15}^{-4/3} \dot{M}_{-3}^{-1/3} P_{\ast,-3}^{-2/3} R_{\ast,6}^{-2/3} \;\si{\s} & \lambda=2\\
                100 \eta^{-3/5} B_{\ast,15}^{-4/5} \dot{M}_{-3}^{-3/5} P_{\ast,-3}^{-2/5} R_{\ast,6}^{-2/5} \;\si{\s}& \lambda=3
            \end{cases},
        \end{equation}
        where $B_{\ast,x}=B_{\ast}/10^{x}\si{\gauss}$, $\dot{M}_{x}=\dot{M}/10^{x}\si{\msun\per\second}$, $P_{\ast,x}=P_{\ast}/10^{x}\si{\s}$, and $R_{\ast,x}=\rast/10^{x}\si{\cm}$.  For sun-like stars, we would expect (for $\lambda=3$)
        \begin{equation}
            \tau_J \approx \num{2e18}\eta^{-3/5}B_{\ast,1}^{-4/5}\dot{M}_{-21}^{-3/5}P_{\ast,6}^{-2/5}R_{\ast,\odot}^{-2/5}\;\si{\s},
        \end{equation}
        and for Hot Jupiters (also for $\lambda=3$),
        \begin{equation}
            \tau_J \approx \num{3e17}\eta^{-3/5}B_{\ast,0}^{-4/5}\dot{M}_{-22}^{-3/5}P_{\ast,5}^{-2/5}R_{\ast,J}^{-2/5}\;\si{\s},
        \end{equation}
        where $R_{\ast,\odot} = \rast/\SI{1}{\rsun}$ and $R_{\ast,J}=\rast/\SI{1}{R_J}$.
        %$\dot{M}$ is a strong function of $c_{T}$ (which, at fixed $B_{\ast}$, sets the ratio of thermal to magnetic energy density and thus the effective $\lambda$), as well as a weak function of $\varOmega_{\ast}$.  In Equation~\eqref{eq:mdot-an}, we found the mass loss rate as a function of $c_{T}$ only; for rapid rotators, the mass loss rate should be enhanced by centrifugal slinging.  We find that an enhancement factor of the form
        %\begin{equation}
        %    \dot{M}_{\rm cen} = \dot{M}_\mathrm{NRNM}f_{\rm cen};\quad f_{\rm cen} = \exp\left[\frac{\varOmega_{\ast}^{2}\rast^{2}}{c_{T}^{2}}\right],\label{eq:mdot-rot}
        %\end{equation}
        %where $\dot{M}_\mathrm{NRNM}$ is the mass loss rate of the non-rotating, non-magnetic (NRNM) case of the same sound speed (Equation~\ref{eq:mdot-an}), to give values of $\dot{M}$ in rough agreement with our simulation results.
        
        This analysis only applies when the asymptotic velocity is dominated by the rotation.  For slow rotators, the Alfv\'en velocity will be of order the sound speed.\footnote{Technically, the velocity should be of the form
        $v_A^2\sim{R_A^2\varOmega^2_\ast + c_T^2}$, but the equations are intractable when the velocity is in this form.}  In this case, the spindown time is nearly independent of rotation rate (as is visually apparent in Figure~\ref{fig:heatmap}c).
        \begin{equation}
            \tau_J \propto \left(\frac{B_\ast\rast^\lambda}{c_T\dot{M}^\lambda }\right)^{\frac{1}{\lambda-1}}.
        \end{equation}
        It should be noted that the mass loss rate is not a constant $\dot{M}=\SI{e-3}{\msun\per\second}$ (in the proto-magnetar case).  As can be seen in Figure~\ref{fig:heatmap}a, it varies by several orders of magnitude across our parameter space, as a strong function of the sound speed and a weaker, but still significant, function of rotation. Following \citet{Metzger2009}, we can describe this as
        \begin{equation}
            \dot{M}\sim \dot{M}_{\rm NRNM}f_{\rm cen}f_{\rm open},
        \end{equation}
        where $\dot{M}_{\rm NRNM}$ is given by Equation~\ref{eq:mdot-an}, and
        \begin{equation}
            f_{\rm cen} \sim \exp\left[\frac{\rast^2\varOmega^2}{c_T^2}\right],\quad f_{\rm open} \sim \frac{\varPhi_{\rm open}}{4\pi\rast^2 B_\ast}.
        \end{equation}
        Furthermore, when comparing to 2D and 3D results, it should be noted that this analysis assumes spherically symmetric outflow, but in a real simulation, density and $\dot{M}$ are a function of $\theta$.  Moreover, in 3D, we must contend with question of tilted magnetospheres.  \citet{Metzger2011} suggest that the open magnetic field flux $\Phi_{\rm open}$ should be corrected by a factor of $\sqrt{1+\sin^2\alpha}$, and that $\dot{E}\propto f^{4/3}_{\rm open}$ in the non-relativistic limit.  Because $\dot{E}\sim\Omega\dot{J}$, we should also have $\dot{J}\propto (1+\sin^2\alpha)^{2/3}$, and thus:
        \begin{equation}
            \tau_J\propto \left(1 + \sin^2\alpha\right)^{-2/3}.
        \end{equation}

    \subsection{Application to Proto-Magnetar Winds}
    
        Immediately after a successful massive star supernova explosion, the hot proto-neutron star drives a neutrino-heated thermal wind into the surrounding medium as it radiates its gravitational binding energy over a timescale $\tau\sim10-100$\,s \citep{Burrows1995,Janka1996}. The wind has been studied as a site for heavy element $r$-process nucleosynthesis (e.g., \citealt{Qian1996,Otsuki2000,Thompson2001,Thompson2018}). For a sufficiently large magnetic fields, the wind will be magnetically-dominated \citep{Thompson2003b,Thompson2005} and with rapid enough rotation, the spindown power carried in the outflow could energize supernovae and potentially produce GRBs \citep{Thompson2004, Bucciantini2006, Metzger2009, Bucciantini2009, Metzger2011}. The model is of direct relevance to the interpretation of SLSNe \citep{Kasen2010,Woosley2010}. Spindown power has also been suggested as a mechanism for normal Type-IIP SNe \citep{Sukhbold2017}. 
        
\begin{figure}
            \centering{}
            \includegraphics[width=\linewidth]{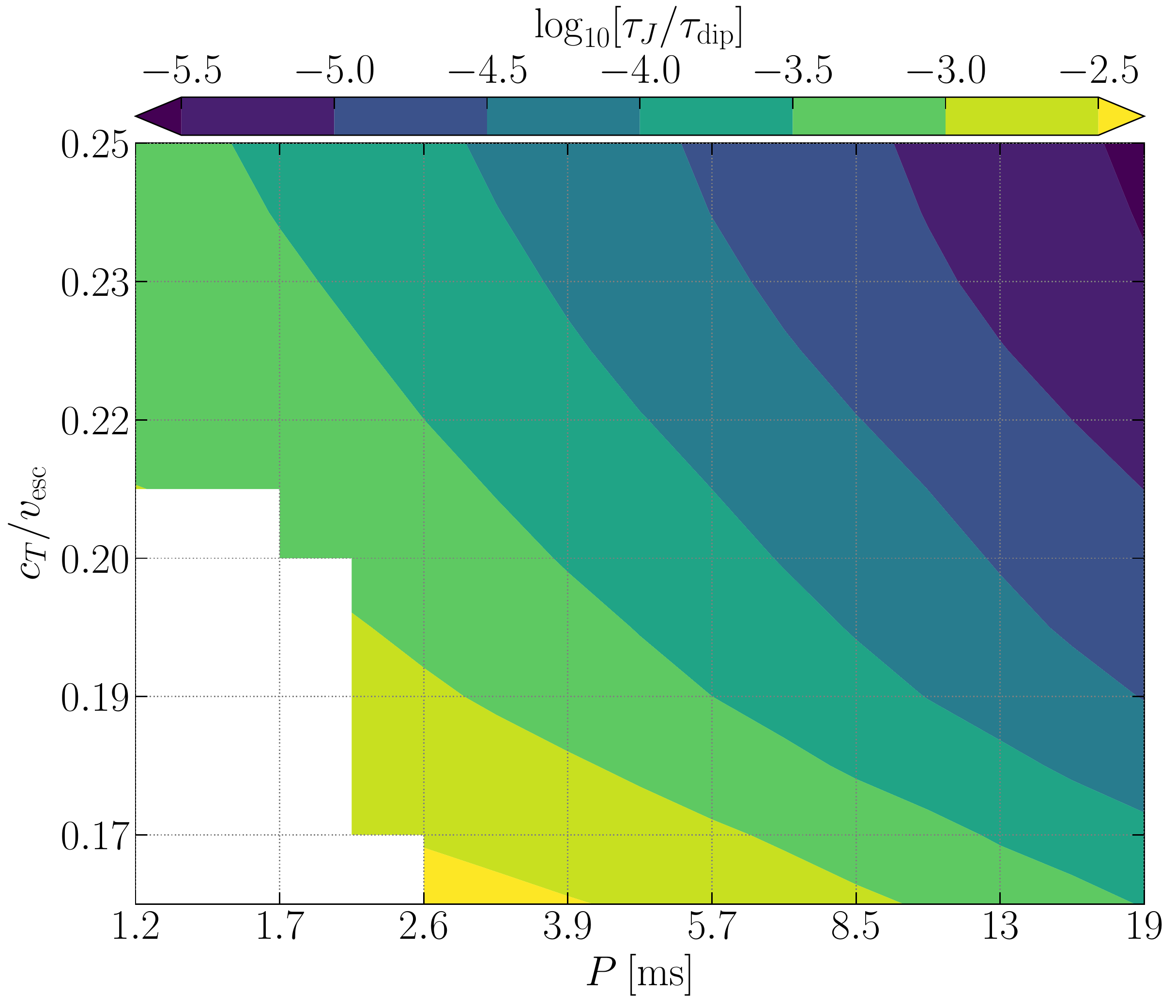}
            \caption{The ratio of the wind spindown time (Equation~\eqref{eq:tauJ}) to the dipole spindown time (Equation~\eqref{eq:taudip}) as a function of sound speed and period, for fixed $B_{\ast}=\SI{e15}{G}$.  For all simulations on our grid, $\tau_{J}\ll\tau_{\rm dip}$.}
            \label{fig:heatmap_spindown}
        \end{figure}

        One metric for assessing when the rotational energy can become dynamically important is when it approaches the kinetic energies of observed explosions
        \begin{equation}
            E_{\rm rot} \simeq \SI{1e51}{\ergs}\,\left(\frac{M_{\ast}}{\SI{1.4}{\msun}}\right)\left(\frac{R_{\ast}}{\SI{12}{\km}}\right)^2\left(\frac{\SI{8}{\ms}}{P}\right)^2.
        \end{equation}
        From fitting numerical models to GRB data and to the observed lightcurves of SLSNe one finds broadly that GRBs require equivalent dipole magnetic field strengths of $B_{\ast}\sim\SI{e15}{\gauss}-\SI{e16}{\gauss}$ \citep{Metzger2011} combined with $P_{\ast}\sim\SI{1}{\ms}-\SI{3}{\ms}$ and for SLSN \citep{Kasen2010,Chatzopoulos2016}, lower $B_{\ast}\sim\SI{e13}{\gauss}-\SI{e14}{\gauss}$ magnetic fields combined with similar spin periods (though some SLSNe are fit by magnetars with magnetic fields approaching $B_{\ast}\sim\SI{e15}{\gauss}$ and longer periods $P_{\ast}\gtrsim\SI{10}{\ms}$, see \citealt{Kasen2010,Hsu2021}). Typically in these models, the spindown is assumed to follow that of a force-free vacuum dipole model:
        \begin{equation}
            \tau_\mathrm{dip} = \frac{E_\mathrm{rot}}{2L_{\rm dip}},\label{eq:taudip}
        \end{equation}
        where $L_{\rm dip}$ is the dipole luminosity:
        \begin{equation}
            L_{\rm dip} = \frac{2R_{\ast}^{6}B_{\ast}^{2}\varOmega_{\ast}^{4}}{3c^{3}} = \frac{2}{3}\sigma\dot{M}\rast^{2}\varOmega_{\ast}^{2}.
        \end{equation}
        This assumption is appropriate after the flow becomes relativistic and Poynting-flux dominated. However, as discussed by \cite{Thompson2004,Metzger2007}, in the early phase of proto-neutron star cooling, the flow is non-relativistic and the spindown timescale is much shorter than a naive application of the dipole expression would indicate. For example, the spindown time as approximated by Equation~\eqref{eq:tauJ-an} for $B_{\ast}=\SI{e15}{G}$, $P_{\ast}=\SI{19}{\milli\second}$, and $\dot{M}=\SI{e-4}{\msun\per\second}$, indicates that the spindown time should be approximately $\tau_{J}\sim\num{6e-4}\:\tau_{\rm dip}$. We show the ratio of the measured spindown time to the dipole spindown time for our full grid of 2D simulations in Figure~\ref{fig:heatmap_spindown}.  As expected, our wind-coupled spindown times are smaller than the dipole spindown times by many orders of magnitude.
        
        One implication of this is that the initial rotation speed of the proto-magnetar must be in excess of the rotation speeds assumed by the dipole spindown models.  Otherwise, the enhanced wind-coupled spindown would reduce the rotation rate to the point where the magnetar could not enhance the late-time luminosity of the supernova.  As an example, we can consider the case of the SLSN ASASSN 15lh \citep{Dong2016}, which is inferred from lightcurve modeling\footnote{Including interactions with the circumstellar medium.} \citep{Chatzopoulos2016,Li2020} to be powered by a magnetar with $B_{\ast}\sim\SI{e13}{\gauss} - \SI{e14}{\gauss}$ and $P_{\ast}\simeq\SI{1}{\ms}$. Such a neutron star would have had a spindown timescale of $\tau_{\rm dip}\sim\SI{10}{\day} - \SI{400}{\day}$, assuming dipole spindown.  Using the analytic scalings developed in the previous section, we can scale our measured spindown times to the lower values $B_{\ast}$ inferred from SLSN models; in doing so we find wind-coupled spindown timescales ranging from $\tau_{J}\sim\SI{29.0}{\second}-\SI{183}{\second}$ at the highest $c_{T}$, to $\tau_{J}\sim\SI{61.5}{\second}-\SI{388}{\second}$ at the lowest $c_{T}$.
        
        Furthermore, our models produce substantial energy loss rates, well in excess of $\SI{e51}{\ergs\per\second}$ for millisecond rotators (see Figure~\ref{fig:heatmap}c).  As the cooling epoch lasts $\SI{1}{\second}-\SI{30}{\second}$ (increasing with decreasing $c_T$), this would contribute a substantial amount of energy to the supernova explosion itself, even for magnetars with spin-down times longer than the cooling epoch.  While scaling $B_{\ast}$ down to the magnitudes associated with SLSNe (\SIrange{e13}{e15}{\gauss}) significantly reduces the wind power, at least for the more magnetic SLSNe models, it does not completely eliminate the total wind energy loss as a significant contribution to the total explosion energy.  We can approximate the energy loss rate as:
        \begin{equation}
            \dot{E}\sim \dot{M}v_{\rm M}^{2}.
        \end{equation}
        For small $B_\ast$, the magnetic field adopts a split-monopole configuration and $\dot{M}$ is nearly independent of $B_\ast$.
        We also have $v_{\rm M}^2\sim\sigma^{2/3}\sim B_{\ast}^{4/3}.$  Thus, the energy loss rate should scale as
        \begin{equation}
            \dot{E}\sim B_{\ast}^{4/3}
        \end{equation}
        Scaling our models by this factor, we find that a ASASSN-15lh-like magnetar with $B_{\ast}\sim\SI{e14}{\gauss}$ and $P_{\ast}\sim\SI{1}{\ms}$ would carry a total wind energy $E_{w}\sim\SI{e50}{\ergs}$ during the cooling epoch, about $0.1$ percent of the total explosion energy \citep{Li2020}.  However, a more magnetized model, such as the magnetar powering DES14C1rhg \citep{Hsu2021} with $B_{\ast}\sim\SI{7e14}{\gauss}$ and $P_{\ast}\sim\SI{13}{\ms}$ can have an energy loss rate as large as $\dot{E}\sim\SI{5e51}{\ergs\per\second}$ (accounting for the $B_{\ast}$ scaling), which even if sustained for \SI{1}{\second} would exceed the kinetic energy of the ejecta.
        
        An important note is that our simulations do not capture all of important physics in the proto-neutron star wind problem. In particular, we do not include neutrino heating/cooling or a detailed electron/positron equation of state (but see \citealt{Prasanna2022}. The mapping between our results and the full problem is thus imperfect.  In addition, as discussed by \cite{Thompson2004,Bucciantini2006,Metzger2009}, the flow rapidly becomes relativistic and Poynting-flux dominated, implying that relativistic calculations are necessary to capture the dynamics as the flow transitions to the relativistic regime.  Such calculations are readily performed in \athena, and will be explored in future papers on this topic.
    
    \subsection{Other Applications}
    
        In addition to proto-magnetar winds, we also scale to regimes representative of Sun-like stars and irradiated hot Jupiters.  Using these scalings, we can apply the simulations discussed in this paper
        %\footnote{We note that the parameter space considered in Figure~\ref{fig:heatmap} and the accompanying section is not always appropriate for the regimes of interest, and we consider simulations outside that space when necessary.}
        to these systems.  We can also apply the analytic estimates developed in the previous section to these other contexts.  See Table~\ref{tab:scaling} and Appendix~\ref{app:units} for more information on our unit scalings.
        
        \subsubsection{Sun-like Stars}\label{sec:solar}
          Low-mass main sequence stars are magnetically active and drive outflows.  Indeed, the original Parker wind model \citep{Parker1958} was developed to describe our own Sun's wind.  Understanding the magnetocentrifugal braking mechanism in these stars is crucial to understanding and modeling the spindown evolution of these stars.  While our models do not capture the more complex field geometries seen in the Sun and similar stars \citep{DeRosa2012,Saikia2016,Finley2017,Finley2018}, they still provide a useful starting point for investigating the physics of these objects (and the dipole component of the field may be the most important for the wind structure and evolution, e.g., \citealt{Finley2017}).
          
          \citet{Finley2017} (hereafter, \citetalias{Finley2017}) consider polytropic wind models of sun-like stars with $\xi_{\varOmega}=\num{6.307e-3}$ (a factor of $\sim\!3$ smaller than the most slowly rotating model in our grid), $\xi_T = 0.25$ (at the surface), and a range of $\xi_{B}$ from \numrange{0.1}{24} (\numrange{2}{500} times larger than our fiducial $\xi_{B}$).  The principal quantity they measure in their simulations is the average Alfv\'en radius (Equation~\ref{eq:meanra}), which they use as a measure of the normalized spindown torque.  They find that this quantity is well-fit by a power law in the quantity
          \begin{equation}
            \varUpsilon_{\rm open} \frac{\langle v_{A}\rangle}{v_{\esc}} \equiv \frac{\varPhi_{\rm open}^{2}}{\dot{M}\rast^{2}v_{\rm esc}}\frac{\langle v_{A}\rangle}{v_{\esc}},
          \end{equation}
          where $\langle v_{A}\rangle$ is the average of the Alfv\'en velocity over all points on $\surf_{A}$.
          
\begin{figure}
            \centering{}
            \includegraphics[width=\linewidth]{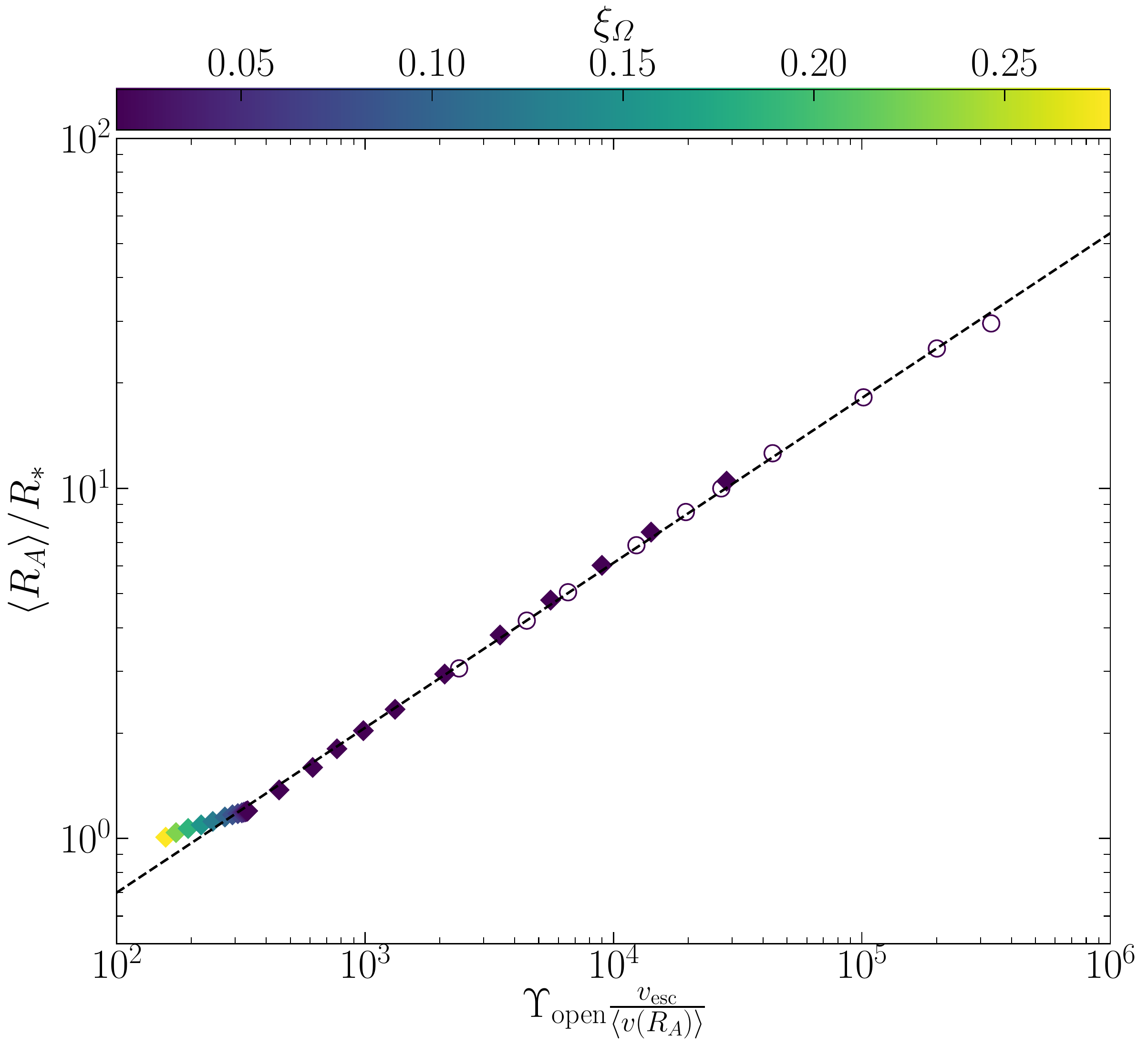}
            \caption{The normalized average Alfv\'en radius $\langle R_{A}\rangle/\rast$ versus the wind magnetization, as measured in our simulations (diamonds) and as reported by \citet{Finley2017} for their pure dipole model (open circles).  Models are colored according to the dimensionless rotation rate $\xi_{\varOmega}$.  The dashed line shows the fit to the \citetalias{Finley2017} data as discussed therein.  We see that our simulations fall along this fit except at the highest rotation rates.}
            \label{fig:avg-ra}
          \end{figure}

          In Figure~\ref{fig:avg-ra}, we plot our measurements of $\langle R_{A}\rangle$ compared to those of \citetalias{Finley2017}.  We compare a collection of slow rotators $\xi_{\varOmega}=\num{6.447e-3}$ at varying $\xi_{B}$, with sound speed $\xi_T=0.247$, to 10 pure dipole simulations from \citetalias{Finley2017}.  In addition, we also show the value of $\langle R_{A}\rangle$ for simulations of more rapid rotators with $\xi_{B}=0.046$.  We find that the slow rotators are well-fit by the empirical fit found in \citetalias{Finley2017}\footnote{Itself following the semi-analytic solution developed in \citet{Pantolmos2017}.}, but the rapid rotators quickly deviate from this relation.  This is in line with the analytic expectations we developed in the previous section.
          
        \subsubsection{Hot Jupiters}
            Hot Jupiters with orbital and spin periods of order $P_\ast\sim\SI{1}{\day}$ have plasma atmospheres ionized by the intense radiation of their host stars.  Models of these atmospheres \citep{MurrayClay2009} (hereafter, \citetalias{MurrayClay2009}) show that they drive outflows that are well modeled by an isothermal, Parker-type wind\footnote{Though their simulations also include tidal gravity.} with $c_{T}\simeq\SI{10}{\km\per\second}$.  These planets should also have strong magnetic fields, though perhaps weaker than Jupiter's field due to their slower rotation period.  Though the influence of their host stars (through their gravity, radiation, and magnetic fields) will have a significant effect on the structure and evolution of these planets' winds, our existing simulations can provide a useful point of comparison for this problem.
            
\begin{figure}
                \centering{}
                \includegraphics[width=\linewidth]{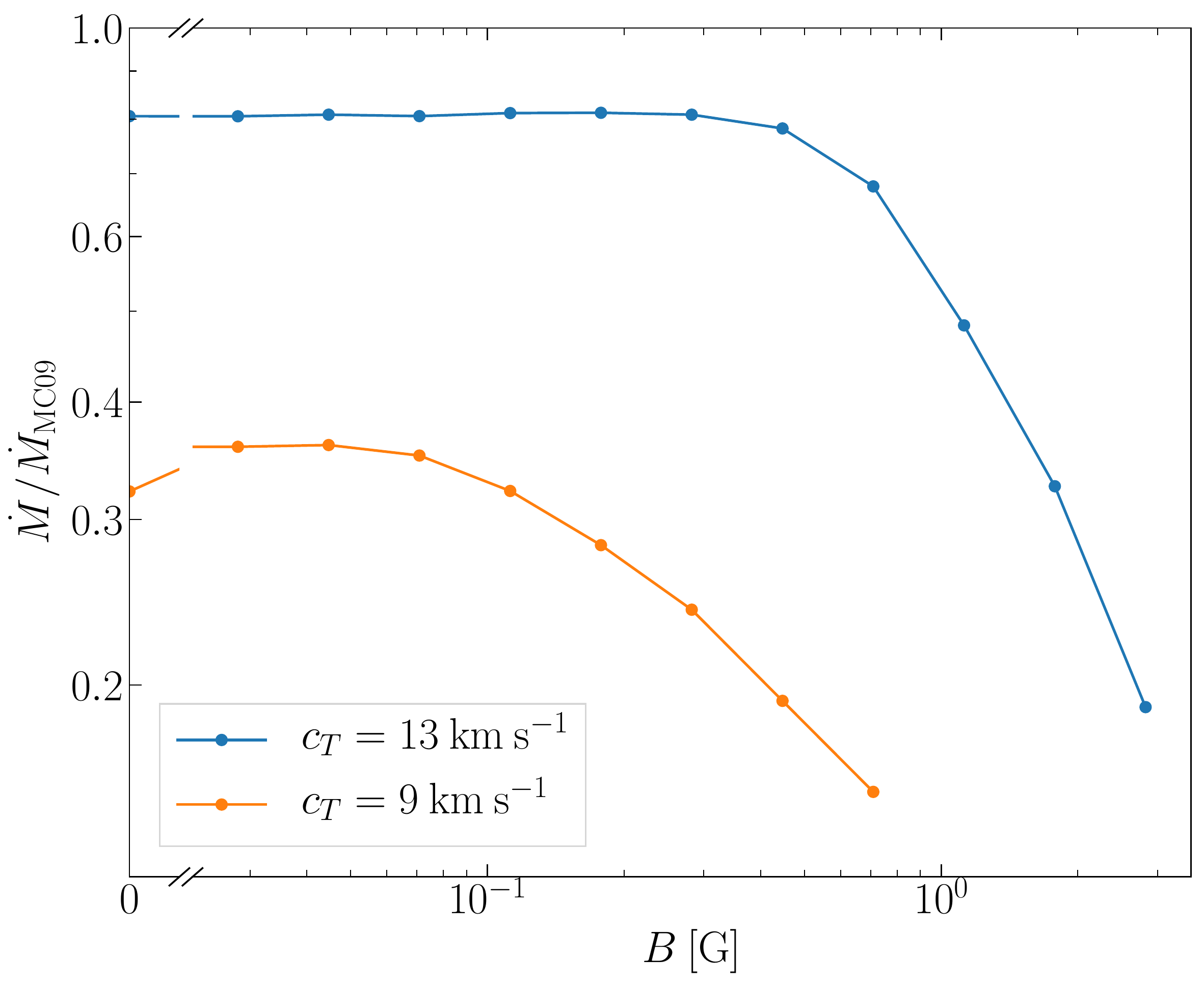}
                \caption{The mass loss rate, relative to the values in \citetalias{MurrayClay2009} measured in our hot Jupiter models as a function of the magnetic field strength $B$.  We show this for a hot (blue) and cool (orange) model, corresponding to the low and high $F_{\rm UV}$ models run by \citetalias{MurrayClay2009}.  $\dot{M}$ approaches a constant for small $\xi_B$ in both cases, with a value a few times smaller than reported by \citetalias{MurrayClay2009}.  We note that in both cases, maximum $\dot{M}$ is achieved at $B>0$; this is a consequence of the non-zero rotation rate of these simulations.} 
                \label{fig:hj-curves}
            \end{figure}

            In Figure~\ref{fig:hj-curves}, we show the measured $\dot{M}$ across a range of magnetic field strengths appropriate to the hot Jupiter regime ($B_\ast~\sim~\SI{1}{\gauss}$), for a warm ($c_T=\SI{13}{\km\per\sec}$) and cool ($c_T=\SI{9}{\km\per\sec}$) model, each with a period of $P_\ast=\SI{96}{\hour}$.  These values correspond to $\xi_B\sim2.442,0.7724$, $\xi_T=0.2341,0.3485$, $\xi_\varOmega=\num{4.876e-2}$ in our dimensionless units.  We find a range of $\dot{M}$, from as low as $\SI{7e9}{\gram\per\second}$ at the largest $B_{\ast}$ considered, to as high as $\SI{7e10}{\gram\per\second}$ at the shortest $P_{\ast}$ considered.  These values are similar to those found by \citetalias{MurrayClay2009} for the lowest incident UV flux (e.g., their case of a hot Jupiter orbiting a sun-like main sequence star, which has $\dot{M}=\SI{3.3e10}{\gram\per\second}$).  For larger UV fluxes, the effective sound speed of the wind increases, which causes the mass loss rate of the wind to increase exponentially (as in Equation~\ref{eq:mdot-an}).  For their example of a hot Jupiter orbiting a T Tauri star, they use a sound speed of $c_T\simeq\SI{30}{\km\per\second}$, and find a mass loss rate 200 times greater than with the main sequence star, (which is roughly the $\dot{M}$ expected analytically for that $c_T$).  It should be noted that these models are 1D and non-magnetic.  At larger $B$ ($B\gtrsim\SIrange{e-1}{1}{\gauss}$, depending on $c_T$), magnetic confinement can significantly reduce the mass loss rate, as can be seen in Figure~\ref{fig:hj-curves}.
            
            \citet{Owen2014} perform 2D simulations including the effects of a magnetic field, but exclusively for high UV fluxes.  Their measured mass loss rates are smaller than those found by \citetalias{MurrayClay2009} by a factor of $\sim\!10$, though still larger than our values by nearly the same factor.  Even were the higher UV fluxes accounted for, however,  the \citet{Owen2014} simulations show that the planet's magnetic field is opened up by the host star's magnetic field, leading to a significant difference in the mass loss rate between the day and night sides of the planet, an effect that cannot be captured in our simulations.
            
            We also measure the spindown rates of the wind.  While hot Jupiters are tidally locked to their host stars and cannot actually spin down, the spindown time can still be used to infer the relative strength of the torque the wind exerts on the planet.  The fastest spindown time we measure is $\tau_{J}=\SI{3.4e9}{yr}$, for extremely fast rotation periods of $P=\SI{14.5}{\hour}$.  For more reasonable rotation periods, the spindown times exceed a Hubble time.  From this, we can infer that the spindown torque caused by the wind is unlikely dynamically important for hot Jupiters, except possibly for anomalously fast rotators early in their evolution (i.e., before they have had time to become tidally locked).  %The spindown torque may be more important in the case of Jupiter mass planets that are not tidally locked to their star; a wind driven by the heat of formation could cause significant spindown in a rapidly rotating Jupiter mass planet over the course of several Gyr.
            
            %Finally, we also consider a case with a tilted magnetosphere $(\alpha=\pi/4)$ for the case with $B_{\ast}=\SI{0.221}{G}$.  We find that in this case, the mass loss rate decreases by 1.7 per cent and the spindown time decreases by 3.8 per cent when compared to the aligned (2D) case at the same $B_{\ast}$.
            
            %\begin{figure*}
            %    \centering{}
            %    \includegraphics[width=\textwidth]{plots/HJ-curves.pdf}
            %    \caption{The mass loss rate (blue) and spindown time (orange) measured in our hot Jupiter models as a function of $B_{\ast}$ (for fixed $P_{\ast}=\SI{85.97}{h}$) and $P_{\ast}$ (for fixed $B_{\ast}=\SI{0.221}{G}$).  The circles in the left-hand panel and the square in the right-hand panel show the mass loss rates measured by \citet{Owen2014} and \citet{MurrayClay2009}, respectively.  The difference in our measured $\dot{M}$ can be partially attributed to the large UV flux used in the case of \citet{Owen2014}, and the lack of a magnetic field in the case of \citet{MurrayClay2009}.}
            %    \label{fig:hj-curves}
            %\end{figure*}
            
    \section{Conclusions}\label{sec:Conclusions}
    
        We have presented results from a suite of non-relativistic two-dimensional MHD simulations exploring the mass loss and spindown rates of stars characterized by a range of rotation rates and sound speeds.  Our simulations show the trends in $\dot{M}$, $\dot{J}$, and $\dot{E}$ with respect to the core rotation rate, isothermal sound speed, and magnetic field strength (Figure~\ref{fig:heatmap}).  We also investigate 3D simulations where the magnetosphere is tilted with respect to the axis of rotation (Figure~\ref{fig:vsalpha_params}).  We also make discuss our simulations in the context of sun-like stars (Figure~\ref{fig:avg-ra}) and Hot Jupiters (Figure~\ref{fig:hj-curves}).  We find our simulations to be in rough agreement with the existing literature on these topics (e.g., \citealt{Finley2017,Finley2018} and \citealt{MurrayClay2009,Owen2014}), though additional work to be done in order to make a more apposite comparison.
        
        There are several targets for improvement we must make in order to improve the accuracy of our simulations in the proto-magnetar context. First, as discussed by \cite{Thompson2004}, the wind velocity at the Alfv\'en point rapidly approaches $c$ during the proto-neutron star cooling epoch and the flow becomes Poynting-flux dominated and relativistic. We also measure wind magnetizations $\sigma>1$ in our most extreme simulations. For this reason, we need to include a treatment of relativistic MHD using the modules in \athena. The other major improvements to be made are additional microphysics, realistic EOS \citep{Coleman2020}, and neutrino heating and cooling.  We have already begun this approach in \citet{Prasanna2022}.  With such improvements in play, we should be able to answer additional questions about the magnetar model of various astrophysical phenomena, including GRBs, SLSNe, and \rproc nucleosynthesis in proto-magnetar winds. \citep{Thompson2018}.
    
    \section*{Acknowledgments}
        We thank Jim Stone, Kengo Tomida, and Adam Finley, and Asif ud-Doula for helpful discussions. TAT thanks Brian Metzger, Niccolo Bucciantini, and Eliot Quataert for discussions and collaboration on this and related topics. TAT acknowledges support from a Simons Foundation Fellowship and an IBM Einstein Fellowship from the Institute for Advanced Study, Princeton.  TAT and MJR are supported in part by NASA grant 80NSSC20K0531. MC acknowledges support from the U.~S.\ Department of Energy Office of Science and the Office of Advanced Scientific Computing Research via the Scientific Discovery through Advanced Computing (SciDAC4) program and Grant DE-SC0018297 (subaward 00009650) and support from the U.~S.\ National Science Foundation (NSF) under Grants AST-1714267 and PHY-1804048 (the latter via the Max-Planck/Princeton Center (MPPC) for Plasma Physics).
    
    \section*{Data Availability}
        The data underlying this article will be shared on reasonable request to the corresponding author.
        
%%%%%%%%%%%%%%%%%%%%%%%%%%%%%%%%%%%%%%%%%%%%%%%%%%
%%%%%%%%%%%%%%%%%%% REFERENCES %%%%%%%%%%%%%%%%%%%
%%%%%%%%%%%%%%%%%%%%%%%%%%%%%%%%%%%%%%%%%%%%%%%%%%
  \bibliographystyle{mnras}
  \bibliography{References/References}

%%%%%%%%%%%%%%%%%%%%%%%%%%%%%%%%%%%%%%%%%%%%%%%%%%
%%%%%%%%%%%%%%%%%%% APPENDICES %%%%%%%%%%%%%%%%%%%
%%%%%%%%%%%%%%%%%%%%%%%%%%%%%%%%%%%%%%%%%%%%%%%%%%
    
    \appendix
    
    \section{Rotating Reference Frame}\label{app:rotatingframe}
        Consider a rotating reference frame, which rotates with respect to the inertial, laboratory frame at a constant angular velocity $\vecomega_{\ast}$.  The coordinates transform between the (primed) rotating frame and the (unprimed) inertial frame according to:
        \begin{align}
            r &\to r'\\
            \theta &\to \theta'\\
            \phi &\to \phi' + \varOmega_{\ast} t'.
        \end{align}
        In addition to the coordinates, the velocity and magnetic fields differ between the reference frames.  They transform as:
        \begin{align}
            \mathbfit{v} &\to \mathbfit{v}' + \vecomega_{\ast}\times\mathbfit{r}'\\
            \mathbfit{B} &\to \mathbfit{B}'.
        \end{align}
        Although only $\phi'$ and $\mathbfit{v}'$ differ from their inertial counterparts, we will still be explicit with our use of primes throughout these appendices to make it clear exactly when we are working in the rotating frame.
        
        \subsection{Coriolis and Centrifugal Forces}
            The motion of a particle in a rotating reference frame behaves as if acted on by the ``fictitious'' Coriolis
            \begin{equation}
                \mathbfit{F}'_\mathrm{cor} \equiv \left.\frac{\partial \rho \mathbfit{v}'}{\partial t'}\right|_\mathrm{cor} = -2\vecomega_{\ast}\times\rho\mathbfit{v}'
            \end{equation}
            and centrifugal
            \begin{equation}
                \mathbfit{F}'_\mathrm{cen} \equiv \left.\frac{\partial \rho \mathbfit{v}'}{\partial t'}\right|_\mathrm{cen} = -\vecomega_{\ast}\times\rho\left(\vecomega_{\ast}\times\mathbfit{r}'\right)
            \end{equation}
            forces.  In the code, these are implemented as additional source terms.  In principle, the centrifugal force is straightforward:
            \begin{equation}
                \mathbfit{F}'_\mathrm{cen} = r'\rho\varOmega_{\ast}^{2}\sin\theta'\left(\uvec{r}\sin\theta' + \uvec{\theta}\cos\theta'\right).
            \end{equation}
            However, to achieve better accuracy, we take the average of the centrifugal force term over the cell:
            \begin{equation}
                \langle\mathbfit{F}'_{\rm cen}\rangle = \frac{\rho\varOmega_\ast^2\partial\phi}{\partial V}\int_{r_i}^{r_{i+1}}\int_{\theta_j}^{\theta_{j+1}}(r')^2\sin\theta'\mathbfit{F}'_\mathrm{cen}\:\mathrm{d}r'\:\mathrm{d}\theta',
            \end{equation}
            where $\partial V$ is the cell volume and $\partial\phi$ is the cell width.  This leads to a velocity source term
            \begin{align}
                \partial\rho\mathbfit{v}'_{\rm cen} &= \langle\mathbfit{F}'_{\rm cen}\rangle\partial t'\\
                \partial\rho\mathbfit{v}'_{\rm cen} &= \varXi
                \left(\begin{array}{c}
                -9\Delta_1[\cos\theta']-\Delta_1[\cos(3\theta')]\\
                4\Delta_3[\sin\theta]\\0
                \end{array}\right)
            \end{align}
            where $\partial t'$ is the timestep,
            \begin{align}
              \varXi &\equiv \frac{\rho\varOmega_{\ast}\Delta_{4}[r]\partial\phi\partial\psi'}{48\partial V},\\
              \partial\psi' &\equiv \varOmega_\ast\partial t',
            \intertext{and we define}
                \Delta_{a}[f] &\equiv f_{i+1}^{a} - f_{i}^{a}
            \end{align}
            for some quantity $f$ across cell faces $i$ and $i+1$.
            
            When implementing the Coriolis force, we must update the momenta\footnote{Though $\rho$ remains constant through this step.} semi-implicitly \citep{Crank1947} in order to conserve energy:
            \begin{equation}
                \rho\mathbfit{v}'_{\rm new} = \partial\psi'\frac{\mathbfit{F}'_{\rm new}+\mathbfit{F}'_{\rm old}}{2} + \rho\mathbfit{v}'_{\rm old},
            \end{equation}
            where $\mathbfit{F}'$ here is the Coriolis force.
            Solving this implicit equation, we find that the Coriolis source term is
            \begin{align}
                \partial\rho\mathbfit{v}'_{\rm cor} &= \frac{\partial\psi'\left(\rho\mathbfit{u}'_{\rm cor} - 2\uvec{z}\times\rho\mathbfit{v}'\right) - (\partial\psi')^2\rho\mathbfit{v}'}{1+(\partial\psi)'^2},\\
                % \rho\mathbfit{v}'_{\rm new} &= \frac{\rho\mathbfit{v}'_{\rm old} + \partial\psi'\left(\rho\mathbfit{v}'_{\rm cor} - 2\uvec{z}\times\rho\mathbfit{v}'_{\rm old}\right)}{1+(\partial\psi)'^2},\\
                \intertext{where}
                \rho\mathbfit{u}'_{\rm cor} &= \rho\partial\psi'\left(\begin{array}{c}
                      v_r'\cos(2\theta') -  v_\theta'\sin(2\theta')\\
                     -v_r'\sin(2\theta') -  v_\theta'\cos(2\theta)\\
                     -v_\phi'
                \end{array}\right).
            \end{align}
            % Specifically, the Coriolis term takes the form
            % \begin{align}
            %     \rho\mathbfit{v}'_{\rm new} &= \frac{\rho\partial\psi'}{1+(\partial\psi')^2}\Big[\big(\partial\psi'(\mathbfit{v}'_{\rm old} + 
            % \end{align}
            % \begin{equation}
            %     \rho\mathbfit{v}'_{\rm new} = \varXi_{2}\left({}
            %     \begin{array}{c}
            %         \psi(v_{r,\rm old}'\cos2\theta' - v_{\theta,\rm old}'\sin2\theta') + 2v_{\phi,\rm old}'\sin\theta\\
            %         -\psi(v_{r,\rm old}'\sin2\theta' - v_{\theta,\rm old}'\cos2\theta') + 2v_{\phi,\rm old}'\cos\theta\\
            %         2(v_{r,\rm old}'\sin\theta' - v_{\theta,\rm old}'\cos\theta') + \varOmega_{\ast}\partial t' v_{\phi,\rm old}'.
            %     \end{array}
            %     \right),
            % \end{equation}
            % where
            % \begin{equation}
            %   \varXi_{2} = \rho\frac{\varOmega_{\ast}\partial t'}{1+(\varOmega_{\ast}\partial t')^{2}}.
            % \end{equation}
    
    \section{Vector Potential for a Tilted Dipole}\label{app:tilt}
        The initial conditions for the magnetic field we impose are that of a tilted dipole.  Since we must specify this by its vector potential, we derive the form of that potential here.  The vector potential of a magnetic dipole is
        \begin{equation}
            \mathbfit{A}' = \frac{\mathbfit{m}'\times\mathbfit{r}'}{r'^3}.
        \end{equation}
        Let the magnetic moment $\mathbfit{m}'$ be a vector of magnitude $m'=\frac{1}{2}B'_{\ast}R^3$, tilted with respect to $\vecomega{}$ by some angle $\alpha$.  By convention, we choose the plane defined by those two vectors to be the $\phi'=0$ plane (in Cartesian coordinates, the $x'$-$z$ plane).  The components of the vector potential are then:
        \begin{equation}
            \mathbfit{A}' = \frac{B'_{\ast}}{2}\frac{R^3}{r'^2}\left(\begin{array}{c}
                 0\\-\sin\alpha\sin\phi'\\\cos\alpha\sin\theta' - \cos\theta'\cos\phi'\sin\alpha
            \end{array}\right)
        \end{equation}
        We choose the magnitude to be defined by $B'_{\ast}/2$ rather than $B'_{\ast}$ so that the magnitude of the magnetic field at the magnetic pole is
        \begin{equation}
            B'_p = \left.\mathbfit{B}'\cdot\mathbfit{B}'\right|_{r'=R,\theta'=\alpha,\phi'=0} = B'_{\ast}.
        \end{equation}
    
    \section{Units}\label{app:units}
        When we initialize our simulations in hydrostatic equilibrium, we choose to normalize the density profile such that $\rho_\ast=G\mast = \rast = 1$.  These three quantities form a basis over the basic kinematic units of length, mass, and time.\footnote{And, in cgs units, electromagnetic quantities are measured in these kinematic units.}  They, in effect, define a set of dimensionless ``code units'' that can be converted back to physical units only after choosing a physical scale, i.e., choosing values of $\rho_\ast$, $G\mast$, and $\rast$ in some physical unit system.  This means that we can freely change the scale of our simulations without repeating them, simply by choosing the appropriate scale factors, which are given by
        \begin{align}
            \mathtt{L} &= \rast\\
            \mathtt{M} &= \rho_\ast\rast^3\\
            \mathtt{T} &= \sqrt{\frac{\rast^3}{G\mast}}.
        \end{align}
        For instance, to convert, e.g., the simulation velocity output into physical units, you would multiply the velocity by a factor of $\mathtt{LT}^{-1} = \sqrt{G\mast/\rast}$, with appropriate $\mast$ and $\rast$ to the problem of interest.  The choice of scales used in this paper are summarized in Table~\ref{tab:scaling}.
    
    \section{Boundary Conditions}\label{app:bc}
        \subsection{Inner Boundary}
            We wish to enforce boundary conditions on the inner boundary such that we maintain
            \begin{equation}
                \frac{\partial B'_{r'}}{\partial t'} = 0
            \end{equation}
            without directly setting the magnetic field components.  We accomplish this as follows.  First, consider, from Maxwell's equations:
            \begin{align}
                \frac{\partial\mathbfit{B}'}{\partial t'} &= -c\nabla'\times\mathbfit{E}'\\
                \frac{\partial B'_{r'}}{\partial t'} &= -\frac{c}{r'\sin\theta'}\left(\frac{\partial}{\partial\theta}\left(E'_{\phi'}\sin\theta'\right) - \frac{\partial E'_{\theta'}}{\partial\phi'}\right).
            \end{align}
            In the special case of \emph{aligned} rotation (i.e., where $\mathbfit{m}'\cdot\vecomega=0$), all $\frac{\partial}{\partial\phi'}$ terms vanish and fixing $E'_{\phi'}=0$ is sufficient to prevent anomalous growth of the magnetic field.  However, once the magnetosphere is tilted with respect to the rotational axis, azimuthal symmetry can no longer be assumed, and we must fix both $E'_{\phi'}=0$ and $E'_{\theta'}=0$ in order to achieve our stated goals.
            
        \subsection{Outer Boundary}
            At the outer boundary, we wish to enforce:
            \begin{equation}
                \dot{M}=\mathrm{const},\quad\rho=\mathrm{const},\quad\mathbfit{L}'=\mathrm{const},\quad\mathbfit{B}'=\mathrm{const},
            \end{equation}
            Where $\boldsymbol{\tau}'$ is the torque. This comprises 7 constraints (${\mathbfit{L}'\cdot\mathbfit{r}'=0}$ by definition, and thus provides no constraint), sufficient to fix the 3 components each of $\mathbfit{v}'$ and $\mathbfit{B}'$, as well as $\rho$.  The constraints on $\rho$ and $\mathbfit{B}'$ are self explanatory, but some work must be done to turn these constraints into direct constraints of the components of $\mathbfit{v}'$.  The constraint on $\dot{M}$ and that on $\rho$ together give us a constraint on $v_r'$:
            \begin{equation}
                \left.v_r'\right|_\mathrm{gh} = \left.v_r'\right|_\mathrm{outer}\left(\frac{r_\mathrm{outer}'}{r_\mathrm{gh}'}\right)^2
            \end{equation}
            where ``gh'' indicates quantities in the ghost zone and ``outer'' indicates quantities in the last outer active cell. From the angular momentum constraint, we obtain:
            \begin{align}
                \left.v_\theta'\right|_\mathrm{gh} &= \left.v_\theta'\right|_\mathrm{outer}\frac{r_\mathrm{outer}'}{r_\mathrm{gh}'}\\
                \left.v_\phi'\right|_\mathrm{gh} &= \left.v_\phi'\right|_\mathrm{outer}\frac{r_\mathrm{outer}'}{r_\mathrm{gh}'} + \varOmega_{\ast}\left(\frac{r_\mathrm{outer}^2 - r_\mathrm{gh}^2}{r_\mathrm{gh}}\right).
            \end{align}

    \section{Full Table of Results}\label{sec:fulltable}
        In this section we present a full table of wind properties for the simulations presented in Section~\ref{sec:Results}.  All simulations presented here have $\xi_B = 0.046$ and values are scaled as:
        \begin{align}
            \dot{M} &\equiv \dot{M}/4\pi\rast^2\rho_0c_T = \dot{M}/\dot{M}_0\\
            \dot{J} &\equiv \dot{J}/\dot{M}_0\rast^2\varOmega_\ast\\
            \dot{E} &\equiv \dot{E}/\dot{M}_0c_T^2\\
            \tau_J &\equiv \tau_J/(\rast c_T^{-1}).
        \end{align}
    
        % \FloatBarrier
        \vfill\eject
        \begin{table}
            \centering
            \begin{tabular}{ccccccc}
\hline$\xi_T$ & $\xi_\varOmega$ & $\log_{10}\dot{M}$ & $\log_{10}\dot{J}$ & $\log_{10}\dot{E}$ & $\log_{10}\tau_J$ & $\Phi/\Phi_0$\\
\hline\hline
% 1
0.247 & 0.278 & -1.4008 & -1.3939 & -0.5419 & 4.3416 & 0.9758\\
      & 0.228 & -1.4700 & -1.4368 & -0.5965 & 4.3845 & 0.9564\\
      & 0.187 & -1.5273 & -1.4703 & -0.6427 & 4.4179 & 0.9481\\
      & 0.153 & -1.5718 & -1.4946 & -0.6789 & 4.4423 & 0.9523\\
      & 0.126 & -1.6050 & -1.5080 & -0.7060 & 4.4556 & 0.9616\\
      & 0.103 & -1.6283 & -1.5058 & -0.7252 & 4.4534 & 0.9898\\
      & 0.085 & -1.6465 & -1.5117 & -0.7402 & 4.4593 & 0.9968\\
      & 0.069 & -1.6598 & -1.5182 & -0.7510 & 4.4658 & 0.9967\\
      & 0.057 & -1.6690 & -1.5224 & -0.7586 & 4.4700 & 0.9979\\
      & 0.047 & -1.6754 & -1.5257 & -0.7639 & 4.4733 & 0.9989\\
      & 0.038 & -1.6797 & -1.5275 & -0.7674 & 4.4751 & 0.9992\\
      & 0.031 & -1.6827 & -1.5284 & -0.7699 & 4.4761 & 0.9990\\
      & 0.026 & -1.6847 & -1.5291 & -0.7716 & 4.4767 & 0.9994\\
      & 0.021 & -1.6861 & -1.5295 & -0.7727 & 4.4771 & 0.9996\\
      & 0.017 & -1.6870 & -1.5297 & -0.7734 & 4.4774 & 0.9998\\
 % 2
0.240 & 0.278 & -1.4983 & -1.4626 & -0.6136 & 4.4103 & 0.9370\\
      & 0.228 & -1.5800 & -1.4863 & -0.6872 & 4.4339 & 0.9196\\
      & 0.187 & -1.6540 & -1.5326 & -0.7480 & 4.4802 & 0.9100\\
      & 0.153 & -1.7108 & -1.5655 & -0.7951 & 4.5131 & 0.9121\\
      & 0.126 & -1.7530 & -1.5815 & -0.8303 & 4.5291 & 0.9244\\
      & 0.103 & -1.7832 & -1.5806 & -0.8556 & 4.5282 & 0.9534\\
      & 0.085 & -1.8060 & -1.5785 & -0.8746 & 4.5261 & 0.9747\\
      & 0.069 & -1.8231 & -1.5780 & -0.8889 & 4.5256 & 0.9870\\
      & 0.057 & -1.8355 & -1.5783 & -0.8993 & 4.5259 & 0.9949\\
      & 0.047 & -1.8443 & -1.5801 & -0.9066 & 4.5277 & 0.9965\\
      & 0.038 & -1.8505 & -1.5818 & -0.9117 & 4.5295 & 0.9974\\
      & 0.031 & -1.8547 & -1.5828 & -0.9152 & 4.5304 & 0.9984\\
      & 0.026 & -1.8576 & -1.5834 & -0.9176 & 4.5311 & 0.9986\\
      & 0.021 & -1.8595 & -1.5838 & -0.9193 & 4.5314 & 0.9989\\
      & 0.017 & -1.8608 & -1.5840 & -0.9204 & 4.5316 & 0.9990\\
% 3
0.232 & 0.278 & -1.5902 & -1.4833 & -0.6932 & 4.4309 & 0.9121\\
      & 0.228 & -1.7000 & -1.5472 & -0.7876 & 4.4948 & 0.8818\\
      & 0.187 & -1.7916 & -1.6074 & -0.8643 & 4.5550 & 0.8710\\
      & 0.153 & -1.8624 & -1.6457 & -0.9240 & 4.5933 & 0.8764\\
      & 0.126 & -1.9135 & -1.6600 & -0.9673 & 4.6076 & 0.8956\\
      & 0.103 & -1.9526 & -1.6640 & -1.0004 & 4.6117 & 0.9165\\
      & 0.085 & -1.9826 & -1.6660 & -1.0258 & 4.6136 & 0.9314\\
      & 0.069 & -2.0048 & -1.6641 & -1.0445 & 4.6118 & 0.9460\\
      & 0.057 & -2.0210 & -1.6618 & -1.0583 & 4.6094 & 0.9561\\
      & 0.047 & -2.0327 & -1.6589 & -1.0681 & 4.6065 & 0.9656\\
      & 0.038 & -2.0410 & -1.6571 & -1.0751 & 4.6047 & 0.9712\\
      & 0.031 & -2.0467 & -1.6550 & -1.0799 & 4.6027 & 0.9762\\
      & 0.026 & -2.0507 & -1.6532 & -1.0832 & 4.6008 & 0.9806\\
      & 0.021 & -2.0534 & -1.6509 & -1.0855 & 4.5986 & 0.9849\\
      & 0.017 & -2.0552 & -1.6497 & -1.0870 & 4.5973 & 0.9879\\
% 4
0.225 & 0.278 & -1.6915 & -1.5248 & -0.7819 & 4.4724 & 0.8616\\
      & 0.228 & -1.8270 & -1.6172 & -0.8957 & 4.5648 & 0.8390\\
      & 0.187 & -1.9400 & -1.6916 & -0.9917 & 4.6392 & 0.8315\\
      & 0.153 & -2.0249 & -1.7334 & -1.0643 & 4.6810 & 0.8401\\
      & 0.126 & -2.0886 & -1.7532 & -1.1190 & 4.7008 & 0.8514\\
      & 0.103 & -2.1385 & -1.7625 & -1.1617 & 4.7101 & 0.8666\\
      & 0.085 & -2.1766 & -1.7638 & -1.1943 & 4.7114 & 0.8814\\
      & 0.069 & -2.2056 & -1.7640 & -1.2191 & 4.7116 & 0.8922\\
      & 0.057 & -2.2274 & -1.7657 & -1.2376 & 4.7133 & 0.8970\\
      & 0.047 & -2.2431 & -1.7664 & -1.2509 & 4.7140 & 0.8995\\
      & 0.038 & -2.2541 & -1.7655 & -1.2603 & 4.7131 & 0.9034\\
      & 0.031 & -2.2618 & -1.7641 & -1.2668 & 4.7117 & 0.9070\\
      & 0.026 & -2.2671 & -1.7629 & -1.2713 & 4.7105 & 0.9098\\
      & 0.021 & -2.2708 & -1.7619 & -1.2744 & 4.7095 & 0.9118\\
      & 0.017 & -2.2733 & -1.7612 & -1.2765 & 4.7088 & 0.9132\\
\hline\end{tabular}

            \caption{Table of wind properties for all simulations considered in Section~\ref{sec:Results}.}
            \label{tab:sims-full}
        \end{table}
        
        \begin{table}
            \centering
            \begin{tabular}{ccccccc}
\hline$\xi_T$ & $\xi_\varOmega$ & $\log_{10}\dot{M}$ & $\log_{10}\dot{J}$ & $\log_{10}\dot{E}$ & $\log_{10}\tau_J$ & $\Phi/\Phi_0$\\
\hline\hline
%5
0.217 & 0.278 & -1.8019 & -1.5857 & -0.8756 & 4.5333 & 0.8184\\
      & 0.228 & -1.9627 & -1.6959 & -1.0129 & 4.6435 & 0.8015\\
      & 0.187 & -2.0957 & -1.7772 & -1.1278 & 4.7248 & 0.7998\\
      & 0.153 & -2.1983 & -1.8289 & -1.2165 & 4.7765 & 0.7998\\
      & 0.126 & -2.2778 & -1.8548 & -1.2853 & 4.8024 & 0.8083\\
      & 0.103 & -2.3408 & -1.8711 & -1.3398 & 4.8187 & 0.8167\\
      & 0.085 & -2.3897 & -1.8782 & -1.3819 & 4.8258 & 0.8228\\
      & 0.069 & -2.4280 & -1.8874 & -1.4148 & 4.8350 & 0.8186\\
      & 0.057 & -2.4564 & -1.8935 & -1.4392 & 4.8411 & 0.8144\\
      & 0.047 & -2.4769 & -1.8957 & -1.4567 & 4.8433 & 0.8135\\
      & 0.038 & -2.4916 & -1.8979 & -1.4692 & 4.8455 & 0.8111\\
      & 0.031 & -2.5020 & -1.8992 & -1.4780 & 4.8468 & 0.8094\\
      & 0.026 & -2.5091 & -1.9003 & -1.4841 & 4.8479 & 0.8078\\
      & 0.021 & -2.5140 & -1.9011 & -1.4883 & 4.8487 & 0.8065\\
      & 0.017 & -2.5174 & -1.9017 & -1.4912 & 4.8493 & 0.8055\\
%6
0.210 & 0.278 & -1.9209 & -1.6556 & -0.9788 & 4.6032 & 0.7845\\
      & 0.228 & -2.1118 & -1.7905 & -1.1446 & 4.7382 & 0.7693\\
      & 0.187 & -2.2638 & -1.8780 & -1.2768 & 4.8256 & 0.7589\\
      & 0.153 & -2.3850 & -1.9346 & -1.3829 & 4.8822 & 0.7644\\
      & 0.126 & -2.4820 & -1.9698 & -1.4677 & 4.9175 & 0.7608\\
      & 0.103 & -2.5603 & -1.9915 & -1.5359 & 4.9391 & 0.7628\\
      & 0.085 & -2.6228 & -2.0049 & -1.5902 & 4.9525 & 0.7604\\
      & 0.069 & -2.6722 & -2.0170 & -1.6328 & 4.9646 & 0.7526\\
      & 0.057 & -2.7097 & -2.0261 & -1.6651 & 4.9737 & 0.7448\\
      & 0.047 & -2.7371 & -2.0304 & -1.6886 & 4.9780 & 0.7402\\
      & 0.038 & -2.7571 & -2.0356 & -1.7057 & 4.9832 & 0.7336\\
      & 0.031 & -2.7713 & -2.0403 & -1.7178 & 4.9879 & 0.7274\\
      & 0.026 & -2.7812 & -2.0460 & -1.7263 & 4.9936 & 0.7200\\
      & 0.021 & -2.7880 & -2.0510 & -1.7321 & 4.9986 & 0.7135\\
      & 0.017 & -2.7927 & -2.0549 & -1.7360 & 5.0025 & 0.7086\\
%7
0.202 & 0.187 & -2.4319 & -1.9712 & -1.4350 & 4.9188 & 0.7389\\
      & 0.153 & -2.5774 & -2.0382 & -1.5623 & 4.9858 & 0.7378\\
      & 0.126 & -2.7009 & -2.0912 & -1.6698 & 5.0388 & 0.7275\\
      & 0.103 & -2.7978 & -2.1175 & -1.7541 & 5.0651 & 0.7230\\
      & 0.085 & -2.8746 & -2.1320 & -1.8193 & 5.0796 & 0.7108\\
      & 0.069 & -2.9378 & -2.1426 & -1.8741 & 5.0903 & 0.7066\\
      & 0.057 & -2.9873 & -2.1503 & -1.9167 & 5.0979 & 0.6994\\
      & 0.047 & -3.0249 & -2.1568 & -1.9489 & 5.1044 & 0.6940\\
      & 0.038 & -3.0522 & -2.1583 & -1.9722 & 5.1059 & 0.6920\\
      & 0.031 & -3.0720 & -2.1607 & -1.9890 & 5.1083 & 0.6882\\
      & 0.026 & -3.0861 & -2.1631 & -2.0009 & 5.1107 & 0.6841\\
      & 0.021 & -3.0958 & -2.1643 & -2.0092 & 5.1119 & 0.6818\\
      & 0.017 & -3.1025 & -2.1635 & -2.0148 & 5.1111 & 0.6821\\
%8
0.194 & 0.153 & -2.7913 & -2.1642 & -1.7616 & 5.1118 & 0.6910\\
      & 0.126 & -2.9336 & -2.2175 & -1.8871 & 5.1651 & 0.6835\\
      & 0.103 & -3.0529 & -2.2534 & -1.9917 & 5.2010 & 0.6694\\
      & 0.085 & -3.1531 & -2.2865 & -2.0769 & 5.2341 & 0.6424\\
      & 0.069 & -3.2334 & -2.3004 & -2.1469 & 5.2480 & 0.6361\\
      & 0.057 & -3.2967 & -2.3045 & -2.2016 & 5.2521 & 0.6357\\
      & 0.047 & -3.3459 & -2.3070 & -2.2437 & 5.2547 & 0.6337\\
      & 0.038 & -3.3829 & -2.3036 & -2.2751 & 5.2513 & 0.6362\\
      & 0.031 & -3.4101 & -2.3014 & -2.2980 & 5.2490 & 0.6381\\
      & 0.026 & -3.4297 & -2.2979 & -2.3144 & 5.2455 & 0.6394\\
      & 0.021 & -3.4436 & -2.2960 & -2.3260 & 5.2436 & 0.6410\\
      & 0.017 & -3.4534 & -2.2912 & -2.3341 & 5.2388 & 0.6428\\
\hline\end{tabular}

            \caption*{(continued)}
        \end{table}
        
        \begin{table}
            \centering
            \begin{tabular}{ccccccc}
\hline$\xi_T$ & $\xi_\varOmega$ & $\log_{10}\dot{M}$ & $\log_{10}\dot{J}$ & $\log_{10}\dot{E}$ & $\log_{10}\tau_J$ & $\Phi/\Phi_0$\\
\hline\hline
%9
0.187 & 0.153 & -3.0154 & -2.2921 & -1.9778 & 5.2397 & 0.6572\\
      & 0.126 & -3.1818 & -2.3518 & -2.1266 & 5.2994 & 0.6388\\
      & 0.103 & -3.3311 & -2.4062 & -2.2582 & 5.3538 & 0.6118\\
      & 0.085 & -3.4549 & -2.4463 & -2.3643 & 5.3939 & 0.5861\\
      & 0.069 & -3.5556 & -2.4601 & -2.4526 & 5.4077 & 0.5830\\
      & 0.057 & -3.6400 & -2.4775 & -2.5246 & 5.4251 & 0.5683\\
      & 0.047 & -3.7053 & -2.4855 & -2.5799 & 5.4332 & 0.5626\\
      & 0.038 & -3.7546 & -2.4819 & -2.6212 & 5.4295 & 0.5644\\
      & 0.031 & -3.7920 & -2.4811 & -2.6521 & 5.4287 & 0.5612\\
      & 0.026 & -3.8195 & -2.4851 & -2.6745 & 5.4328 & 0.5545\\
      & 0.021 & -3.8392 & -2.4899 & -2.6903 & 5.4375 & 0.5473\\
      & 0.017 & -3.8529 & -2.4913 & -2.7013 & 5.4390 & 0.5443\\
%a
0.179 
%~ & 0.278 & -2.4555 & -1.9830 & -1.5104 & 4.9306 & 0.8906\\
      %~ & 0.228 & -2.7401 & -0.5064 & 1.2505 & 3.4540 & 7.7365\\
      %~ & 0.187 & -2.9683 & -2.1018 & -1.6299 & 5.0494 & 1.7042\\
      & 0.153 & -3.2555 & -2.4330 & -2.2185 & 5.3806 & 0.6145\\
      & 0.126 & -3.4635 & -2.5247 & -2.4065 & 5.4723 & 0.5729\\
      & 0.103 & -3.6463 & -2.5974 & -2.5693 & 5.5450 & 0.5380\\
      & 0.085 & -3.8020 & -2.6629 & -2.7023 & 5.6105 & 0.4982\\
      & 0.069 & -3.9330 & -2.7091 & -2.8149 & 5.6567 & 0.4711\\
      & 0.057 & -4.0366 & -2.7324 & -2.9030 & 5.6800 & 0.4566\\
      & 0.047 & -4.1173 & -2.7413 & -2.9706 & 5.6889 & 0.4481\\
      & 0.038 & -4.1793 & -2.7352 & -3.0222 & 5.6828 & 0.4491\\
      & 0.031 & -4.2283 & -2.7409 & -3.0617 & 5.6885 & 0.4428\\
      & 0.026 & -4.2649 & -2.7511 & -3.0905 & 5.6987 & 0.4333\\
      & 0.021 & -4.2909 & -2.7465 & -3.1110 & 5.6941 & 0.4321\\
      & 0.017 & -4.3098 & -2.7496 & -3.1258 & 5.6973 & 0.4284\\
%b
0.172 
%~ & 0.278 & -2.5697 & -2.0533 & -1.6274 & 5.0009 & 0.8776\\
      %~ & 0.187 & -3.1699 & -2.3306 & -2.2086 & 5.2782 & 0.7695\\
      %~ & 0.153 & -3.4601 & -2.4840 & -2.4768 & 5.4316 & 0.6772\\
      & 0.126 & -3.7759 & -2.7147 & -2.7369 & 5.6623 & 0.5047\\
      & 0.103 & -3.9985 & -2.8054 & -2.9390 & 5.7531 & 0.4628\\
      & 0.085 & -4.1891 & -2.8879 & -3.1026 & 5.8356 & 0.4246\\
      & 0.069 & -4.3525 & -2.9470 & -3.2409 & 5.8946 & 0.3903\\
      & 0.057 & -4.4890 & -3.0051 & -3.3516 & 5.9527 & 0.3632\\
      & 0.047 & -4.5967 & -3.0478 & -3.4369 & 5.9954 & 0.3376\\
      & 0.038 & -4.6756 & -3.0705 & -3.4983 & 6.0181 & 0.3287\\
      & 0.031 & -4.7373 & -3.0891 & -3.5456 & 6.0367 & 0.3168\\
      & 0.026 & -4.7788 & -3.0734 & -3.5777 & 6.0210 & 0.3192\\
      & 0.021 & -4.8118 & -3.0704 & -3.6023 & 6.0180 & 0.3149\\
      & 0.017 & -4.8377 & -3.0876 & -3.6213 & 6.0352 & 0.3044\\
%c
0.164 
%~ & 0.228 & -3.0905 & -2.2984 & -1.8560 & 5.2460 & 1.4012\\
      %~ & 0.187 & -3.4114 & -2.4577 & -2.5336 & 5.4053 & 0.7450\\
      %~ & 0.126 & -4.1229 & -2.8974 & -3.1711 & 5.8450 & 0.4630\\
      %~ & 0.103 & -4.3962 & -3.0088 & -3.4336 & 5.9564 & 0.4180\\
      & 0.085 & -4.6305 & -3.1035 & -3.6377 & 6.0511 & 0.3761\\
      & 0.069 & -4.8374 & -3.1924 & -3.8040 & 6.1400 & 0.3436\\
      & 0.057 & -5.0101 & -3.2651 & -3.9304 & 6.2127 & 0.3119\\
      & 0.047 & -5.1437 & -3.3119 & -4.0220 & 6.2595 & 0.2863\\
      & 0.038 & -5.2491 & -3.3484 & -4.0912 & 6.2960 & 0.2669\\
      & 0.031 & -5.3365 & -3.3888 & -4.1479 & 6.3364 & 0.2502\\
      & 0.026 & -5.3972 & -3.4134 & -4.1853 & 6.3611 & 0.2401\\
      & 0.021 & -5.4495 & -3.4452 & -4.2195 & 6.3928 & 0.2263\\
      & 0.017 & -5.4668 & -3.4253 & -4.2253 & 6.3729 & 0.2259\\
\hline\end{tabular}

            \caption*{(continued)}
        \end{table}
    
\FloatBarrier
\bsp{}
\label{lastpage}
\end{document}